\documentclass[twocolumn]{aastex62}

\newcommand{\msun}{{M$_\odot$}}
\newcommand{\Al}{$^{26}$Al}
\newcommand{\Fe}{$^{60}$Fe}
\newcommand{\Ne}{$^{22}$Ne}
\newcommand{\F}{$^{19}$F}
\newcommand{\Cl}{$^{36}$Cl}
\newcommand{\Ca}{$^{41}$Ca}

\usepackage{longtable}

\received{23 March 2023}
\revised{12 May 2023}
\accepted{15 May 2023}
\submitjournal{ApJ}

\shorttitle{Nucleosynthesis Yields from Massive Binary Stars}
\shortauthors{Brinkman et al.}

\begin{document}
\title{ALUMINIUM-26 FROM MASSIVE BINARY STARS III. BINARY STARS UP TO CORE-COLLAPSE AND THEIR IMPACT ON THE EARLY SOLAR SYSTEM }

\correspondingauthor{Hannah Brinkman}
\email{hannah.brinkman@kuleuven.be}

\author{Hannah E. Brinkman}
\affil{Konkoly Observatory, Research Centre for Astronomy and Earth Sciences (CSFK), E\"otv\"os Lor\'and Research Network (ELKH), MTA Centre of Excellence, Konkoly Thege Mikl\'os \'ut 15-17, H-1121 Budapest, Hungary}
\affiliation{Graduate School of Physics, University of Szeged, Dom t\'er 9, Szeged, 6720 Hungary}
\affiliation{Institute of Astronomy, KU Leuven, Celestijnenlaan 200D, 3001, Leuven, Belgium}

\author{Carolyn Doherty}
\affiliation{School of Physics and Astronomy,
Monash University, VIC 3800, Australia}

\author{Marco Pignatari}
\affil{Konkoly Observatory, Research Centre for Astronomy and Earth Sciences (CSFK), E\"otv\"os Lor\'and Research Network (ELKH), MTA Centre of Excellence, Konkoly Thege Mikl\'os \'ut 15-17, H-1121 Budapest, Hungary}
\affil{E. A. Milne Centre for Astrophysics, University of Hull, Hull HU6 7RX, UK}
\affil{NuGrid Collaboration, \url{http://nugridstars.org}}

\author{Onno Pols}
\affil{Department of Astrophysics/IMAPP, Radboud University, P.O. Box 9010, 6500 GL Nijmegen, The Netherlands}

\author{Maria Lugaro}
\affil{Konkoly Observatory, Research Centre for Astronomy and Earth Sciences (CSFK), E\"otv\"os Lor\'and Research Network (ELKH), MTA Centre of Excellence, Konkoly Thege Mikl\'os \'ut 15-17, H-1121 Budapest, Hungary}
\affiliation{School of Physics and Astronomy,
Monash University, VIC 3800, Australia}
\affiliation{ELTE E\"{o}tv\"{o}s Lor\'and University, Institute of Physics, Budapest 1117, P\'azm\'any P\'eter s\'et\'any 1/A, Hungary}

\begin{abstract}
Many of the short-lived radioactive nuclei that were present in the early Solar System can be produced in massive stars. In the first paper in this series \citep{Brinkman1}, we focused on the production of $^{26}$Al in massive binaries. In our second paper \citep{Brinkman2021}, we considered rotating single stars, two more short-lived radioactive nuclei, $^{36}$Cl and $^{41}$Ca, and the comparison to the early Solar System data. In this work, we update our previous conclusions by further considering the impact of binary interactions. We used the MESA stellar evolution code with an extended nuclear network to compute massive (10-80 M$ _{\odot} $), binary stars at various initial periods and solar metallicity (Z=0.014), up to the onset of core collapse. The early Solar System abundances of $^{26}$Al and $^{41}$Ca can be matched self-consistently by models with initial masses $\geq$25 \msun{}, while models with initial primary masses $\geq$35 \msun{} can also match $^{36}$Cl. Almost none of the models provide positive net yields for \F{}, while for \Ne{} the net yields are positive from 30 \msun{} and higher. This leads to an increase by a factor of approximately 4 in the amount of \Ne{} produced by a stellar population of binary stars, relative to single stars. Also, besides the impact on the stellar yields, our 10 \msun{} primary star undergoing Case A mass-transfer ends its life as a white dwarf instead of as a core-collapse supernova. This demonstrates that binary interactions can also strongly impact the evolution of stars close to the supernova boundary. 
\end{abstract}


\section{Introduction} \label{intro}
The presence of radioactive isotopes in the early Solar System (ESS) is well-established as their abundances are inferred from meteoritic data showing excesses in their daughter nuclei. In this work, as in the previous works in this series of papers, the main focus is on \Al{}, a short-lived radioactive isotope (SLR) with a half life of 0.72 Myr \citep[][]{Alhalflife}. We also consider \Cl{} and \Ca{}, with half lives 0.301 Myr \citep[][]{Clhalflife} and 0.0994 Myr \citep[][]{Cahalflife}, respectively. These three radioactive isotopes represent the fingerprint of the local nucleosynthesis that occurred nearby at the time and place of the birth of the Sun. Therefore, they give us clues about the environment and the circumstances of such birth \citep{adams10, Lugaro2018}.\\
\indent Massive star winds have been suggested as a favoured site not only of origin of the \Al{} in the ESS  \citep[see, e.g.,][]{Arnould1997,Arnould2006,Gounelle2012,Gaidos2009,Young2014}, but also of \Cl{} and \Ca{}. This is because these three isotopes can be synthesised in massive stars, and expelled both by their winds and/or due to binary interactions \cite{BraunandLanger,Brinkman1, Brinkman2021} and, in equal or larger amounts, by their final core-collapse supernova \citep{MeyerClayton2000,Lawson2022}. The \Al{} present in the stellar winds is produced by proton captures on $^{25}$Mg during hydrogen burning. The majority of the \Al{} is expelled together with $^{36}$Cl and $^{41}$Ca, which are produced instead during helium burning by neutron captures on the stable isotopes, $^{35}$Cl and $^{40}$Ca, respectively.\\
\indent In \cite{Brinkman1} (hereafter Paper I), we investigated the impact of \textit{binary interactions} on the yields of \Al{} from the primary star (i.e., the initially most massive star) of a binary system. We found, in agreement with \cite{BraunandLanger}, that for initial primary masses up to $\sim$40-45 \msun{}, binary interactions can significantly increase the yields of \Al{}, especially for the lowest masses, 10-25 \msun{}. For these systems, the increase can be as high as a factor of $\sim$150 at 10 \msun{} and a factor $\sim$5 at 25 \msun{}. Above 40-45 \msun{}, the binary interactions do not have an impact on the yield, due to the strong mass loss through winds of these stars, which is comparable to the mass lost due to binary interactions. In this first paper, however, we did not consider \Cl{} and \Ca{}.\\
\indent In \cite{Brinkman2021} (hereafter Paper II), we investigated the yields of the SLRs \Al{}, \Cl, and \Ca{} for \textit{single massive stars}, both rotating and non-rotating, and the impact of these yields on the ESS. We found that stars with initial mass of 40-45 \msun{}, depending on their initial rotational velocity, can explain the presence of \Al{} and \Ca{} in the ESS, but only stars with initial masses of 60 \msun{} and higher can also explain the presence of \Cl{}.\\
\indent In this paper, we explore the effect of the removal of the hydrogen-rich envelope due to binary interactions on the yields of isotopes produced in later burning stages, primarily helium burning. As mentioned above for \Al{}, the impact of binarity is negligible above $\sim$40-45 \msun{}. It is however not clear yet whether for \Cl{} and \Ca{} the binary interactions might have an impact for higher masses than this limit, since these two isotopes are produced in a later stage of the evolution than \Al{}. We thus consider several binary configurations in the mass interval 10-60 M$_{\odot}$ for the primary stars, since for these systems the impact of binary interactions is most prominent, based on the results of Paper I. We also consider one binary model for primary stars of 70 and 80 \msun{}.\\
\indent As in Paper II, we consider the stable isotopes \F{} and \Ne{} to establish the impact of the binary interactions on their yields. \cite{MeynetArnouldF192000} have shown that Wolf-Rayet stars can contribute significantly to the galactic $^{19}$F abundance, while \cite{PalaciosF192005} found that Wolf-Rayets are unlikely to be the source of galactic $^{19}$F, when including updated mass-loss prescriptions and reaction rates. In Paper II, we found that only the most massive stars in our sample produce positive net yields of \F{}. Binary interactions might increase the range for which positive net yields are found. As for $^{22}$Ne, there are puzzling observations of an anomalous $^{22}$Ne/$^{20}$Ne ratio in cosmic rays, a factor of $\sim$5 higher than in the solar wind \citep{Prantzos2012CR}. Comparing models and observations might be a key for finding the source of these cosmic rays, in relation to massive stars and binary systems, as has been done for single stars by \cite{Tatischeff2021Ne22}.\\
\indent The structure of this paper is as follows: in Section \ref{method}, we briefly describe the method and the important input parameters for our models. In Section \ref{Results1} we show the results of the stellar evolution of our models, and make a comparison between the models of Paper II and representative systems of Case A and Case B mass transfer, where Case A mass-transfer occurs during the main-sequence evolution, while Case B mass transfer occurs during the time between core hydrogen and helium burning. In Section \ref{Results2}, we discuss the stellar yields, and again compare the models of Paper II to representative systems of the binary interactions. In Section \ref{sec:Discussion} we discuss our results, put them into the context of the early Solar System (ESS), and consider the composition of the winds of the models that could represent the ESS. We end our paper with the conclusion in Section \ref{sec:Conclusion}.
\section{Method and input physics} \label{method}
As in Paper I and II, we have used version 10398 of the MESA stellar evolution code \citep{MESA1,MESA2,MESA3,MESA4} to calculate massive star models, both single and in a binary configuration. We have included an extended nuclear network of 209 isotopes within MESA such that the stellar evolution and the detailed nucleosynthesis are solved simultaneously. Below we briefly describe the input physics for the single massive stars and the input parameters for the binary systems. Only the key input parameters and the changes compared to the input physics of Papers I and II are discussed. The inlist files used for the simulations are available on Zenodo under a Creative Commons 4.0 license: \url{https://doi.org/10.5281/zenodo.7956527}\\
\indent The physical input parameters for the stellar models presented here are the same as for Paper II. The initial masses of our primary models are 10, 15, 20, 25, 30, 35, 40, 45, 50, 60, 70, and 80 \msun{}. The initial composition used is solar with Z=0.014, following \cite{Asplund2009}. For the initial helium content we have used Y=0.28. Our nuclear network contains all the relevant isotopes for the main burning cycles (H, He, C, Ne, O, and Si) to follow the evolution of the star in detail up to core collapse. All relevant isotopes connected to the production and destruction of $^{26} $Al, $^{36}$Cl, $^{41}$Ca, $^{19}$F, $^{22}$Ne, and $^{60}$Fe are also included into our network. Including the ground and isomeric states of $^{26} $Al, the total nuclear network therefore contains 209 isotopes (see Paper II). Following \cite{Farmer2016} (and references therein) a nuclear network of 204 isotopes is optimal for the full evolution of a star, especially because it includes isotopes that influence the electron fraction, $Y_{\rm e}$, which is important for the core collapse \citep[see][]{Heger2000}.\\
\indent As in the two previous papers, we have used the Ledoux criterion to establish the location of the convective boundaries. The semi-convection parameter, $ \alpha_{sc} $, was set to 0.1 and the mixing length parameter, $ \alpha_{mlt} $, to 1.5. We make use of overshooting via the ``step-overshoot" scheme with $ \alpha_{ov} $=0.2 for the central burning stages. We do not use overshoot on the helium burning shell and the later burning shells. The overshoot on the hydrogen shell was set to $ \alpha_{ov} $=0.1.\\
\indent The mass-loss scheme is the same as in Paper II. For the hot phase (T$_{\rm eff}\geq$ 11 000 K), we use the prescription given by \cite{Vink2000, Vink2001} and for the cold phase (T$_{\rm eff}\leq$ 10000 K) we use \cite{NieuwenhuijzendeJager1990}. For the WR-phase we use \cite{NugisLamers2000}. All phases of the wind have a metallicity dependence $\dot{M}\,\propto\,$Z$^{0.85}$ following \cite{Vink2000} and \cite{VinkdeKoter2005}.\\
\indent We have evolved the stars to the onset of core collapse, using an (iron-)core infall velocity of 300 km/s as the termination point of our simulation, or until a total number of models of 10$^{4}$, since in some cases the cores of the binary models are too small to form an iron-core (see Section \ref{finalstates}).\\
\indent The main focus of this work is on the yields from the non-rotating primary star of the binary system. The binary input is the same as in Paper I, where we used a fixed initial mass-ratio of q=$\frac{M_2}{M_1}$=0.9 and fully non-conservative mass-transfer. With this choice of mass transfer the primary will not accrete any of the mass of the secondary, and all the mass is lost from the system. The range of periods is the same as in Paper I, and based on the stellar radius of the primary. We have selected periods such that mass transfer first occurs either during hydrogen burning, commonly referred to as Case A mass transfer, or after hydrogen burning, but before the central ignition of helium, commonly referred to as Case B mass transfer \citep{KippenhahnWeigert1967}. The chosen periods cover the range of orbital sizes where Case A or Case B mass transfer is initiated while the star still has a radiative envelope (between $~$2 and $~$100 days). Compared to Paper 1, we added a binary system with a primary mass of 10 \msun{} at a period of 104.6 days, as well as a binary system undergoing Case B mass-transfer for both 70 and 80 \msun{}.\\
\indent We used the same stopping criteria for our primary stars as in Paper II, where we evolved the stellar models to the onset of core collapse, or to 10$^{4}$ models. In addition to these two criteria, like in Paper I, we stopped the simulations when reverse mass-transfer starts taking place (R$_{2} \geq$ R$_{L,2}$, where R$_{2}$ is the radius of the secondary star, and R$_{L,2}$ the radius of the Roche lobe of the secondary). At this point, we checked manually whether a common envelope was formed (R$_{1} \geq$ R$_{L,1}$ as well), and if so, we terminated the run since the outcome of such systems is uncertain. If the primary was still within its Roche lobe (R$_{1} <$ R$_{L,1}$), we split the binary and evolved the primary star further as if it were a single star, up to the onset of core collapse, or to 10$^{4}$ models. Because the reverse mass-transfer occurs at the timescale of the secondary star, the binaries are uncoupled at different stages in the evolution of the primary. However, this way, we kept the binary intact for as long as possible. The uncoupling procedure might overlook some effects of binary interactions, such as further mass-loss or mass-gain due to future mass-transfer phases, which have the potential to change the yields significantly compared to the assumptions used here. However, since we use fully non-conservative mass-transfer and no accretion is involved, there is no mass-gain in our models, and only extra mass-loss due to mass-transfer is overlooked. Despite these shortcomings, this procedure is a step closer to producing full binary yields for the isotopes of interest. A similar procedure has been used in recent papers on the effects of binary interactions on carbon yields \citep{Farmer2021} and the structure of the pre-supernova of primary stars \citep{Laplace2021}.
\subsection{Yield calculations}
In this work our focus is on the pre-supernova nucleosynthetic yields from the winds and the mass transfer of the binary systems. To calculate these yields, we integrate over time the surface mass fraction multiplied by the mass loss, because the wind mass loss and the mass transfer are not instantaneous processes, though the mass transfer has a shorter timescale than the wind loss. For the stable isotopes, there are two yields to consider, the total yield and the net yield. The total yield is calculated as described above. The net yield is the total yield minus the initial total mass present in the star. For the SLRs the net yield is identical to the total yield, because the initial mass present in the stars is zero for these isotopes, and thus no distinction will be made for these yields. Hereafter, the yield will always refer to the total yield, unless otherwise indicated.\\
\indent We do not make a distinction between the yields ejected by the stellar wind or by the binary mass-transfer, because, as shown in Paper I, the main effect of the mass transfer is not to increase the yields directly, but to expose the deeper layers of the stars such that the subsequent winds can carry off the isotopes produced in the stellar interior.
\section{Stellar Models: results and discussion}\label{Results1}
\begin{figure*}
    \centering
    \includegraphics[width=\linewidth]{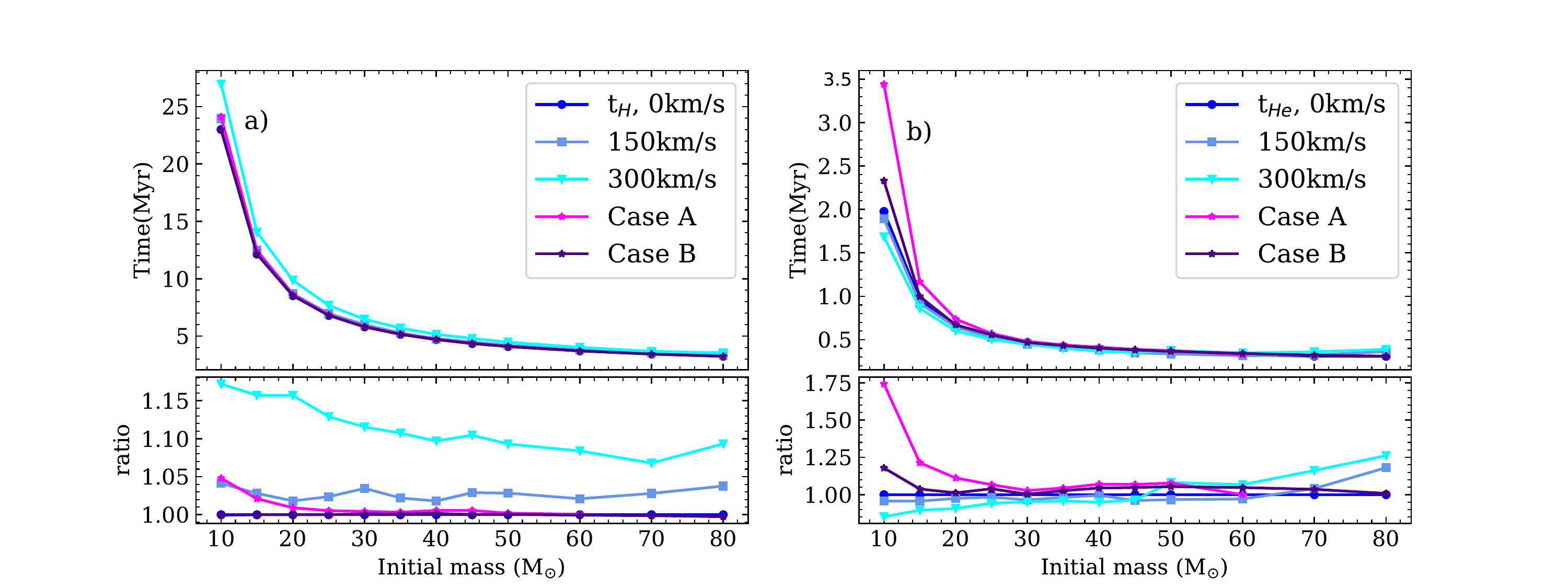}
    \caption{The duration of core hydrogen burning (t$_{\rm H}$, left panel) and core helium burning (t$_{\rm He}$, right panel) for rotating and non-rotating single stars, as well as for representative systems of Case A and Case B mass transfer, as a function of the initial stellar (primary) mass. To highlight the differences, the bottom panels show the ratio between the non-rotating single star model (reference) and the other stars with the same initial mass.}
    \label{Burning}
\end{figure*}
\begin{figure*}
    \centering
    \includegraphics[width=\linewidth]{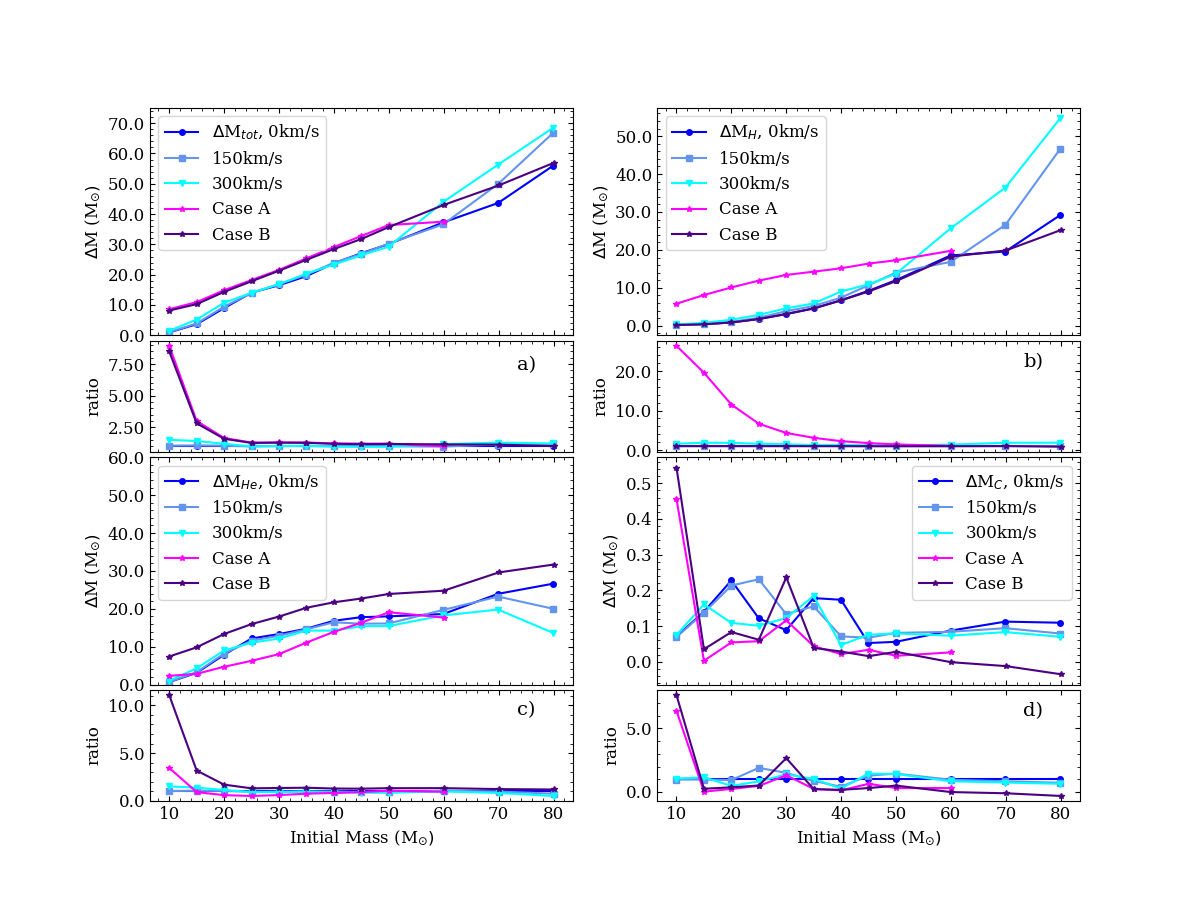}
    \caption{The mass lost in \msun{} from the different models as a function of the initial stellar (primary) mass during the entire evolution of the star ($\Delta$M = M$_{\rm ini}$-M$_{\rm *, f}$, panel a) , the two most important burning phases for the models considered here, hydrogen burning ($\Delta$M$_{\rm H}$ = M$_{\rm ini}$-M$_{*, \rm H}$, panel b), end of hydrogen burning to the end of helium burning} ($\Delta$M$_{\rm He}$ = M$_{*,\rm H}$-M$_{*,\rm He}$, panel c), and the mass loss from the end of helium burning, during carbon burning and beyond ($\Delta$M$_{\rm C}$ = M$_{*,\rm He}$-M$_{*,\rm f}$, panel d). The single stars are indicated by dots, the representative binary systems with stars (indicated in Table \ref{StellarInfo}). The lower panels show the ratio between the non-rotating single star (reference model) and the other stars with the same initial mass.
    \label{MassRatio}
\end{figure*}
\begin{figure}
    \centering
    \includegraphics[width=\linewidth]{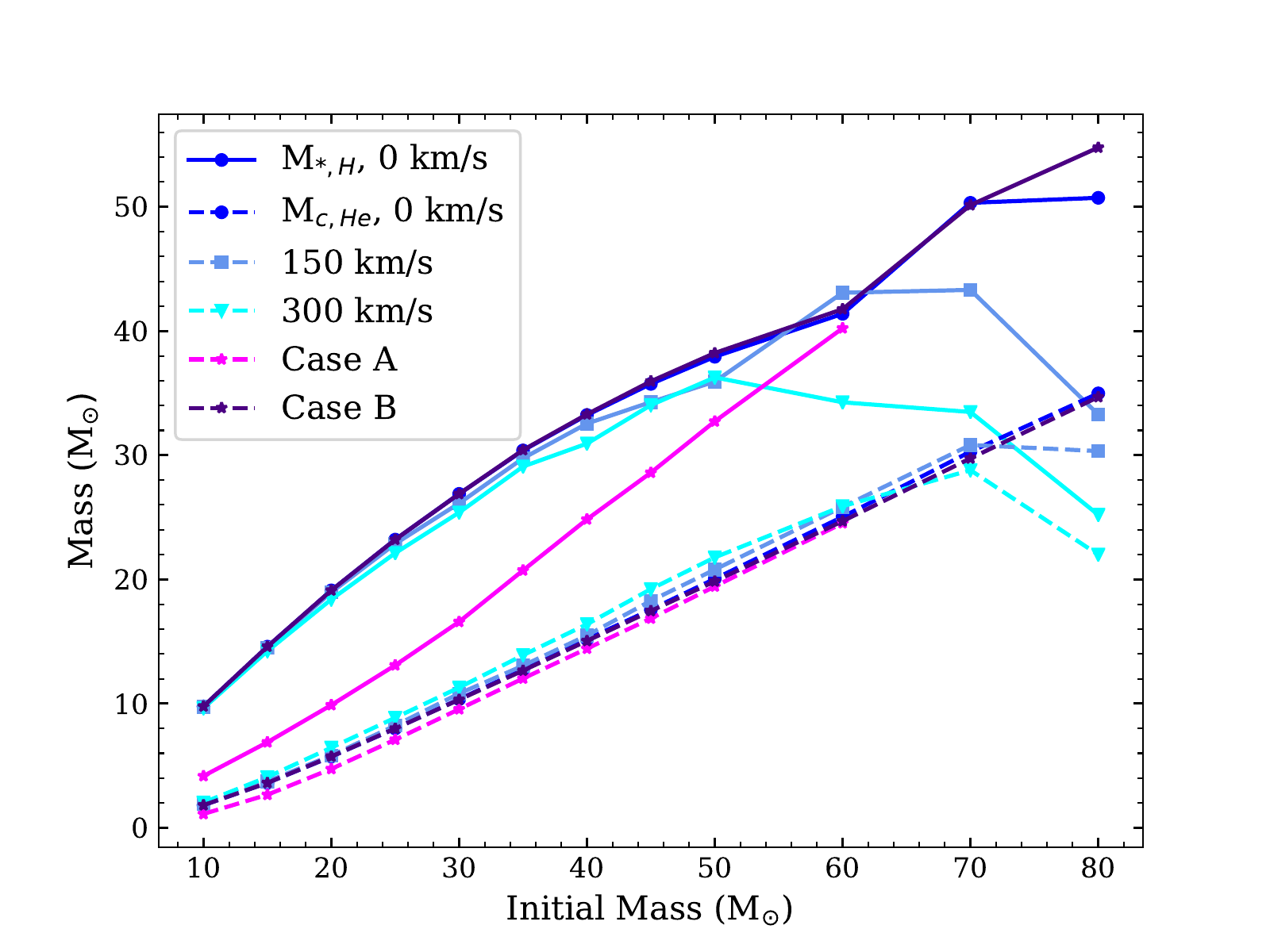}
    \caption{The total stellar mass M$_{*,\rm H}$, (solid lines) and the the hydrogen depleted core mass M$_{c,\rm He}$, (dashed lines) at the end of the main sequence as a function of the initial mass for the three single star models of different initial rotational velocities and for representative binary systems (indicated in Table \ref{StellarInfo}).}
    \label{CoreMass}
\end{figure}
In this section, we discuss the results of the stellar evolution of the binary models, and compare them to the single stars of Paper II, both rotating and non-rotating. Table \ref{StellarInfo} (see Appendix \ref{StellarEvolTable}) contains the key information about the evolutionary stages of the primary stars, as well as the same information from the single, non-rotating star with the same initial masses presented in Paper II. For the 30, 35, 40, 45, and 60 \msun{} models presented here, the system with the shortest period enters a common-envelope phase. Because the assumed spherical symmetry is broken, we did not compute the outcome of these systems with MESA, as it is beyond the scope of this work. Under certain assumptions, however, it is possible to do such calculations, see, e.g., \cite{Marchant2021CEs}. A few other systems did not reach the final stages because of computational issues. In the end, 53 of our 58 models reached a final stage for which we could compute the wind yields.\\
\indent After comparing the models, we briefly consider the effect of binarity on the final stages of the stars, especially for those stars close to the supernova mass-boundary, which is 8-10 \msun{} for single stars (depending on the initial conditions of the star, see, e.g., \citealt{Doherty2017}), as well as the effect of changing the wind prescription for the two lowest masses considered here.
\subsection{The effects of binary interactions and stellar rotation}
\indent In Figures \ref{Burning}-\ref{CoreMass}, a selection of properties of the stellar models are shown for the single star models at different initial rotational velocities (from Paper II), as well as for representative binary systems for Case A and for Case B mass transfer. For the details on the rotating models, see Paper II.\\
\indent As a reminder, in Section 3 of Paper II, we considered the effects of rotation on the evolution, on the total stellar mass, on the mass loss in the various phases of the evolution, on the mass of the helium core, and on the duration of the hydrogen- and helium-burning phase. We showed there that the main effect of the rotational mixing is to extend the lifetime of hydrogen burning, especially of the models rotating at an initial rotational velocity of 300 km/s. For helium burning, rotation has the opposite effect and shortens the duration of this burning phase, except for the initially most massive models. This is shown by the blue lines in Figure \ref{Burning}. Also, the rotating models lose more mass overall than the non-rotating models (see Figure \ref{MassRatio}), which is due to a combination of the extended lifetimes and the rotational boost on the stellar winds.\\
\indent In Figures \ref{Burning} and \ref{MassRatio}, we compare not only non-rotating models with the rotating models, but we also consider the effects of Case A and Case B binary mass-transfer. For each mass, where possible, we have selected representative binary models, undergoing Case A and Case B mass-transfer, indicated by bold-faced periods in Tables \ref{StellarInfo} and \ref{BinYields}. The selected Case A system for 20 \msun{} did not reach core collapse, but ended its evolution during silicon burning. The mass loss for this star is complete, and therefore it is still a representative system. For the 70 and 80 \msun{} models, it was not possible to create a system undergoing Case A mass-transfer without a common envelope. This is because, with an initial mass ratio of 0.9, the hydrogen burning lifetimes of the primary and the secondary become more and more similar as the initial mass of the primary star goes up. This means that for the 70 and 80 \msun{} systems, the primary and the secondary evolve at such a similar pace that they move off the main sequence at almost the same time, which inhibits creating a Case A mass transfer without a common envelope. For Case B, the period is longer and the orbit is wider, and then the slight time difference in the evolution prevents the common envelope from forming. Thus, for 70 and 80 \msun{} we have only a single Case B system each.\\
\indent In Figure \ref{Burning}, the duration of core hydrogen and core helium burning are shown, as well as the ratio of the different models as compared to the single, non-rotating model, which we use as a standard. The Case A binary systems have a slightly extended main sequence phase due to the shrinking of the hydrogen burning core during the mass transfer (see also Appendices A and B of Paper I). This extension is comparable to that of the slower rotational velocity (150 km/s) on the lower mass end, though for the single stars the lifetime is extended due to receiving more fuel rather than a shrinking core as for the Case A binaries. For the higher initial masses, the duration of hydrogen burning for the Case A systems is very similar to the non-rotating single star case. The Case B binary systems, as expected, do not differ from the non-rotating single star case, since the interaction between the two stars only takes place after the main sequence has already finished, and the star has evolved as if it were single up to this point.\\
\indent For core helium-burning, the Case A binaries have the longest lifetimes. This is a direct result of the mass transfer during hydrogen burning. The binary interaction in this phase leads to smaller hydrogen-depleted cores at the end of core hydrogen-burning (see also Figure \ref{CoreMass}). The hydrogen-depleted core is defined as the part of the star where the hydrogen content is below 0.01 and the helium content is above 0.1. Smaller hydrogen-depleted cores have lower helium burning luminosities and therefore longer lifetimes. The Case B systems also have a slightly longer helium-burning lifetime, which is due to the the smaller total masses at the end of helium burning as a result of the mass transfer just before helium-burning starts. These stars have overall smaller helium-depleted cores than their non-interacting counterparts.\\
\indent Figure \ref{MassRatio} shows the total mass loss and the mass loss in different phases of the evolution. In panel a, the total mass loss over the whole evolution is shown, while panels b, c, and d show the mass loss during hydrogen burning, between the end of hydrogen burning and the end of helium burning, and the end of helium burning to the end of the evolution, respectively. Up to 50 \msun{}, the Case A systems lose more mass than the non-rotating single star, while for the Case B systems it is not till 80 \msun{} that the mass loss becomes similar. We remind that for 70 and 80 \msun{} we do not have a binary system undergoing Case A mass transfer. Especially on the lower mass end, up to $\sim$30 \msun{}, the binaries lose significantly more mass than the single stars, even compared to the rotating models. The rotating models lose more mass at the higher mass end, from 60 \msun{} and higher, compared to both the non-rotating model as well as the binaries (for more details on binary evolution versus single star evolution, see the appendix of Paper I).\\
\indent As expected, the Case A systems lose most mass during the early phases of the evolution, though between 50-60 \msun{}, the fastest rotating model becomes the star that loses most mass. The Case B models lose roughly the same amount of mass as the single non-rotating stars, except for the 80 \msun{} model, which is due to the strong winds at this mass. Then, again as expected, the Case B systems lose most mass between the end of hydrogen burning and the end of helium burning, while most of the Case A systems lose less mass than the non-rotating models. Finally, the mass-loss in the final phases (panel d) does not show a predictable behaviour.\\
\indent Figure \ref{CoreMass} shows the total stellar mass and mass of the hydrogen depleted core at the end of hydrogen burning. As expected, the Case A systems have the lowest total masses, as they lose most mass during this phase due to the mass transfer. Their hydrogen depleted cores are slightly smaller than those of the other models. For masses above 60 \msun{}, the fastest rotating models lose most mass in this phase, due to the stronger winds. However, their hydrogen depleted cores are slightly larger than most of the other models, except for the most massive ones, due to the additional internal mixing due to rotation. The Case B systems are nearly identical to the non-rotating single stars.\\
\indent These figures show that the binary interactions not only have an effect on the mass loss of the models, but also on their internal structure. This is in qualitative agreement with the results of \cite{Laplace2021}, who investigated the isotopic distribution in the core of Case B systems, and how this would affect their supernova explosions.
\subsection{Final fates}\label{finalstates}
\begin{figure}
    \includegraphics[width=0.5\textwidth]{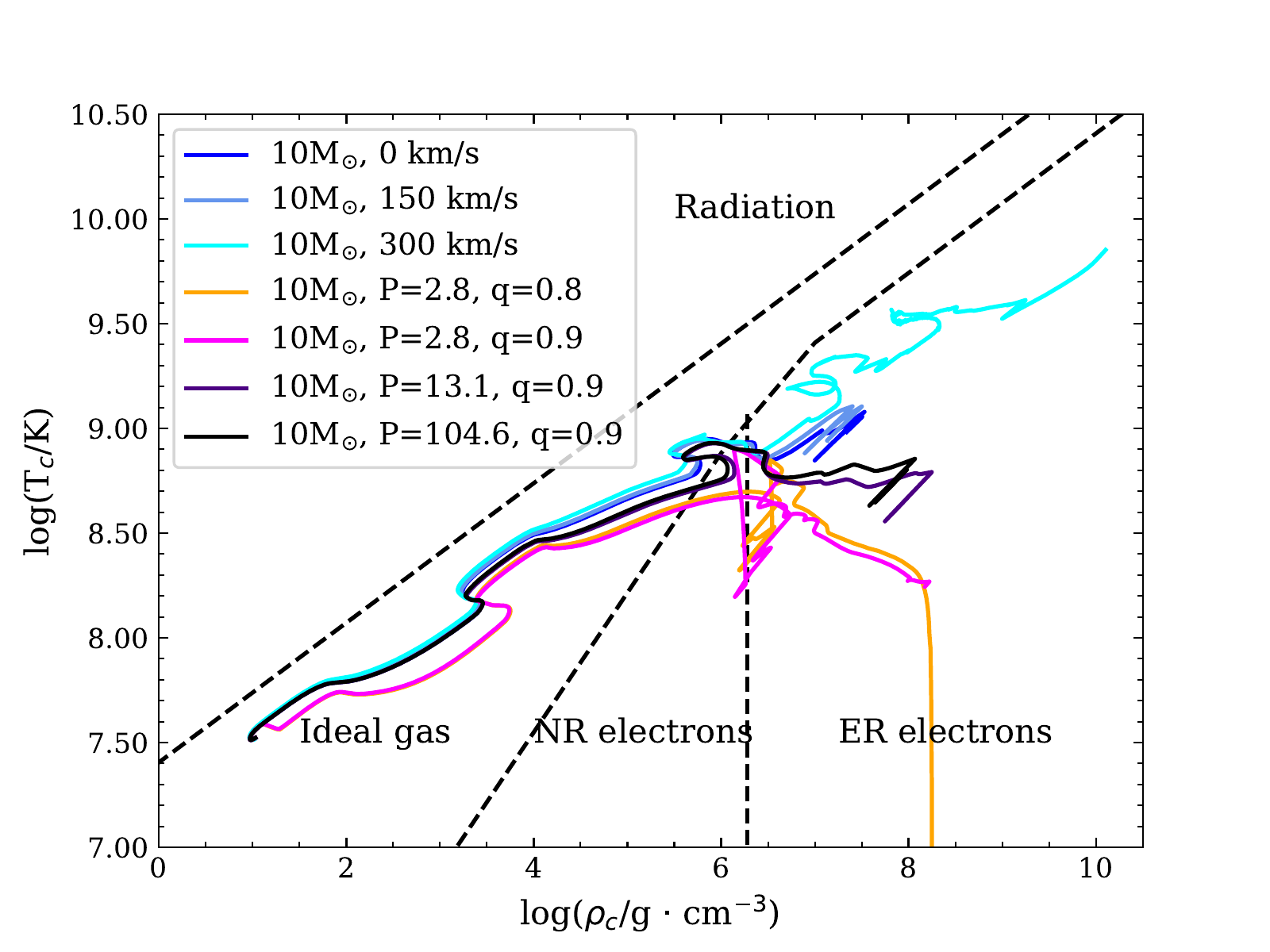}
    \caption{$\rho_{\rm c}$-T$_{\rm c}$ diagram for different 10 \msun{} models. The blue lines are the single star models from Paper II. The magenta, indigo, and black lines are for a 10 \msun{} primary at a period of 2.8, 13.1, and 104.6 days, respectively. The orange line is for a 10 \msun{} primary at a period of 2.8 days, but with a binary mass ratio of q=0.8 instead of q=0.9. The black dashed lines give a rough indication of the equations of state, i.e., radiative, ideal gas, non-relativistic electron pressure (NR electrons), and extremely relativistic electron pressure (ER electrons).}
    \label{fig:rhoctc}
\end{figure}

The most important difference between the single stars and the primary stars of the binary systems is the amount of mass loss, which can lead to significant alteration in the stars of the binary system \citep[see also][]{LangerReview, Laplace2021}. While the majority of the primary stars simulated here end their lives as a core-collapse supernova, the binary systems with an initial primary mass of 10 \msun{} are an exception. The clearest example is the primary star of the system with an initial period of 2.8 days, which is plotted as the magenta line in Figure \ref{fig:rhoctc}. The decline in the central temperature and the central density indicates that this star will end its life as a white dwarf. Eventually, the interior of the star will start cooling at a constant core density, as is shown by the orange line, representing a 10 \msun{} star at a period of 2.8 days, but with a mass ratio of 0.8 instead of 0.9. The other two 10 \msun{} primaries shown, with periods of 13.1 and 104.6 days respectively, end their lives as supernovae, but likely as electron-capture supernovae instead of iron core-collapse supernovae based on their central density and temperature profile \citep[see, e.g., Figure 1 of][]{Tauris2015}. Based on previous studies of binary systems in this mass-range, \citet[][]{ElectronCaptureSNPodsiadlowski2004,Poelarends2017,Siess2018ECSNe}, we expected to find such stars in our models. For comparison also the single stars of Paper II are shown in Figure \ref{finalstates} as blue lines, where the cyan line represents the 10 \msun{} model at 300 km/s. This star shows the clearest signature of becoming an iron core-collapse supernova. This shows that besides the impact on the stellar yields as determined in Paper I and further explored in Section \ref{Results2}, the binary interactions also have a strong impact on the final fate of stars close to the supernova boundary.
\section{Yields from non-rotating binary stars up until core collapse}\label{Results2}
As we showed in Paper I, binary interactions can have a large impact on the yields of massive stars by stripping off the envelope and exposing the deeper layers of the star, especially on the lower mass end of the massive star regime. So far, we only considered the impact of the binary interactions on \Al{}. For the models presented in Paper II, we considered four more isotopes and the impact of rotation, but did not include the impact of binary interactions. In this section, we discuss how binary interactions impact the yields of four isotopes: \Cl{}, \Ca{}, \Ne{}, and \F{}. We also briefly reconsider \Al{}, and compare to the models of Papers I and II. The complete set of wind yields for all isotopes and models presented here, are available on Zenodo under a Creative Commons 4.0432 license: \url{https://doi.org/10.5281/zenodo.7956513}\\
\indent In this section, we use the following definitions:
\begin{itemize}
    \item the ``effective binary yield'' is the single star yield multiplied by the binary enhancement factor, which is defined as the arithmetic average increase (over all periods considered) of the yield of the binary systems compared to the yield of the single stars. We have given all systems an equal weight in the calculations, especially since the yields are not that sensitive to the period in order of magnitude (aside from the models that break down), with the exception of the 10 \msun{} models. We do not consider the effect of higher order systems, such as triples \footnote{This a slightly different definition than what was used in Paper I, where the effective binary yield was already corrected for the binary fraction, which has been renamed the ``effective stellar yield''. For more details see Chapter 5 of \cite{ThesisHannah}}.
    \item the ``maximum binary yield'' is the maximum value of the individual wind yields for the binary systems with a certain initial primary mass,
    \item and the ``minimum binary yield'' is the minimum value of the individual wind yields for the binary systems with a certain initial primary mass. This minimum binary yield includes systems that did not complete their evolution, for example, due to the formation of a common envelope or numerical issues.
\end{itemize}
\subsection{Short-lived radioactive isotopes}
In this section, we discuss the effects of the binary interactions on \Al{}, \Cl{}, and \Ca{}.
\subsubsection{Aluminium-26}
\begin{figure}
    \includegraphics[width=\linewidth]{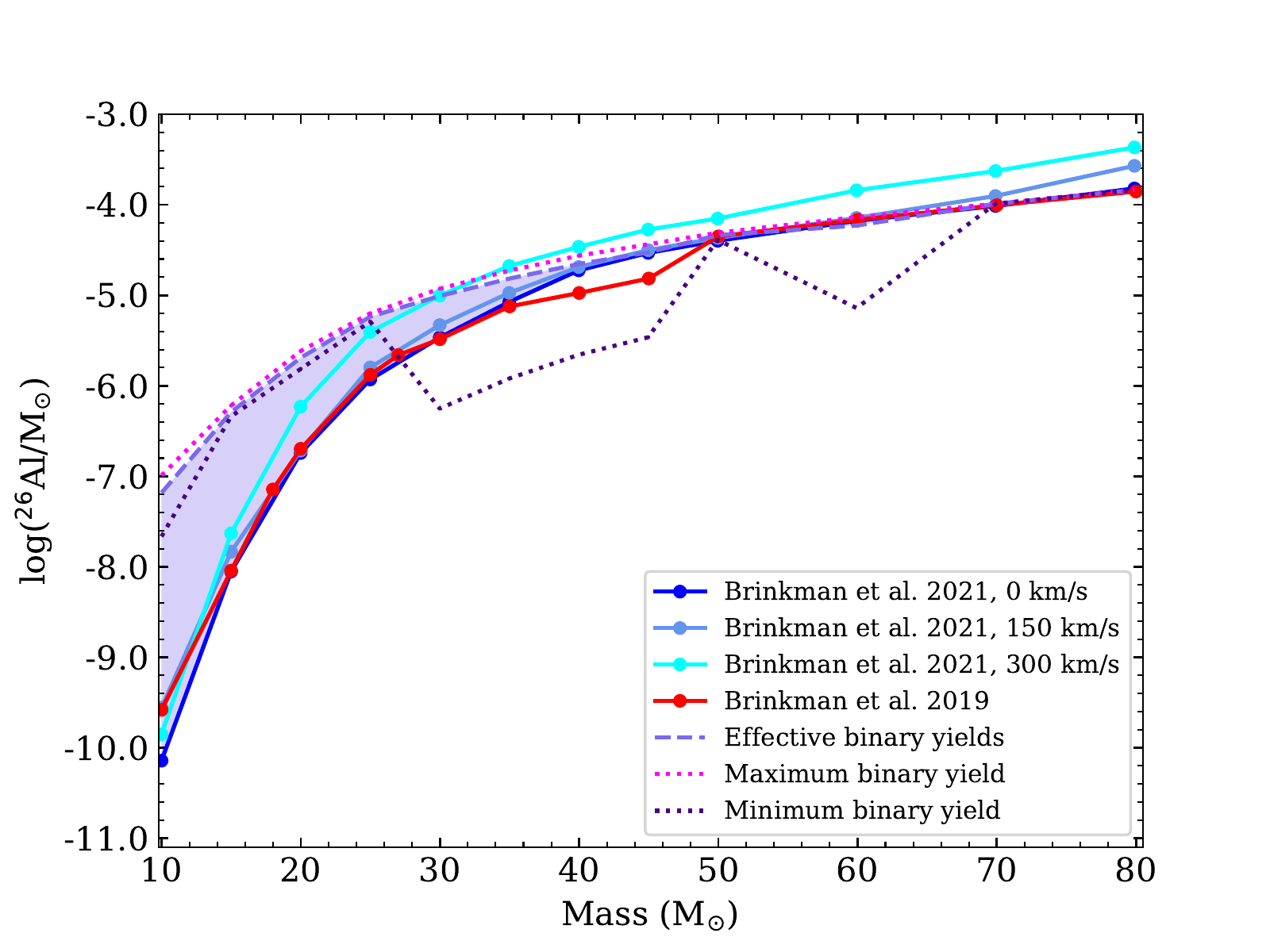}
    \caption{Single star yields for Paper I (red dots) and Paper II (blue dots). The yields are compared to the effective, the maximum, and minimum binary yield for the models presented in this work for \Al{}. The yields for the individual binary systems are given in Table \ref{BinYields}. The shaded area indicates the potential effect of binarity, showing yields between the effective binary yield and the single star yield. This is because the fully non-conservative mass-transfer gives an upper limit for the binary yields.}
    \label{Set4Al}
\end{figure}

Evolving either single or binary stars models up to core-collapse (this work) instead of to the onset of carbon burning (as in Paper I) has a very limited impact on the \Al{} yield. The biggest change is for the 10 \msun{} star, where the model from Paper I has a 3.6 times higher yield than the model from Paper II. The model in Paper I loses slightly more mass than the model in Paper II, 1.01 \msun versus 0.96 \msun{}, for reasons explained in Section 2 of Paper II. At this specific mass such a relatively small decrease in the mass lost from the star is enough to make a significant difference for the \Al{} yield. The 40 and 45 \msun{} models of Paper II have double the yields compared to the yields of Paper I, because those models did not reach the onset of carbon burning in Paper I. The other single star yields are within $\pm$10\% of the yields in Paper I.\\
\indent Figure \ref{Set4Al} shows the effective binary yields for the binary models computed for this work, as well as the maximum binary yield and the minimum binary yield, as defined above, along with the single star yields of Paper I and II. For most initial masses the minimum and maximum binary yields (and hence effective values) are very similar. However, models with masses between 30-45 \msun{} and 60 \msun{} and the shortest initial period enter a common envelope phase that truncates the evolution which leads to substantially lower (minimum) binary yield. As these short period models are only a small subset, the effective binary yield more closely resembles the maximum yield binary predictions.\\
\indent Figure \ref{Set4Al} shows that both rotation and binary mass transfer increase the yields of \Al{}. At $\sim$ 30 \msun{} the fastest rotating model of Paper II gives a similar yield to the binary system, while this happens only around $\sim$ 40 \msun{} for the lower initial rotational velocity, and around 50 \msun{} for the non-rotating models. This demonstrates that for higher rotational velocities, the increase in the wind mass loss washes out the effect of binary interactions at a lower mass than for the non-rotating models.
\subsubsection{Chlorine-36 and Calcium-41}
\begin{figure*}
    \includegraphics[width=\textwidth]{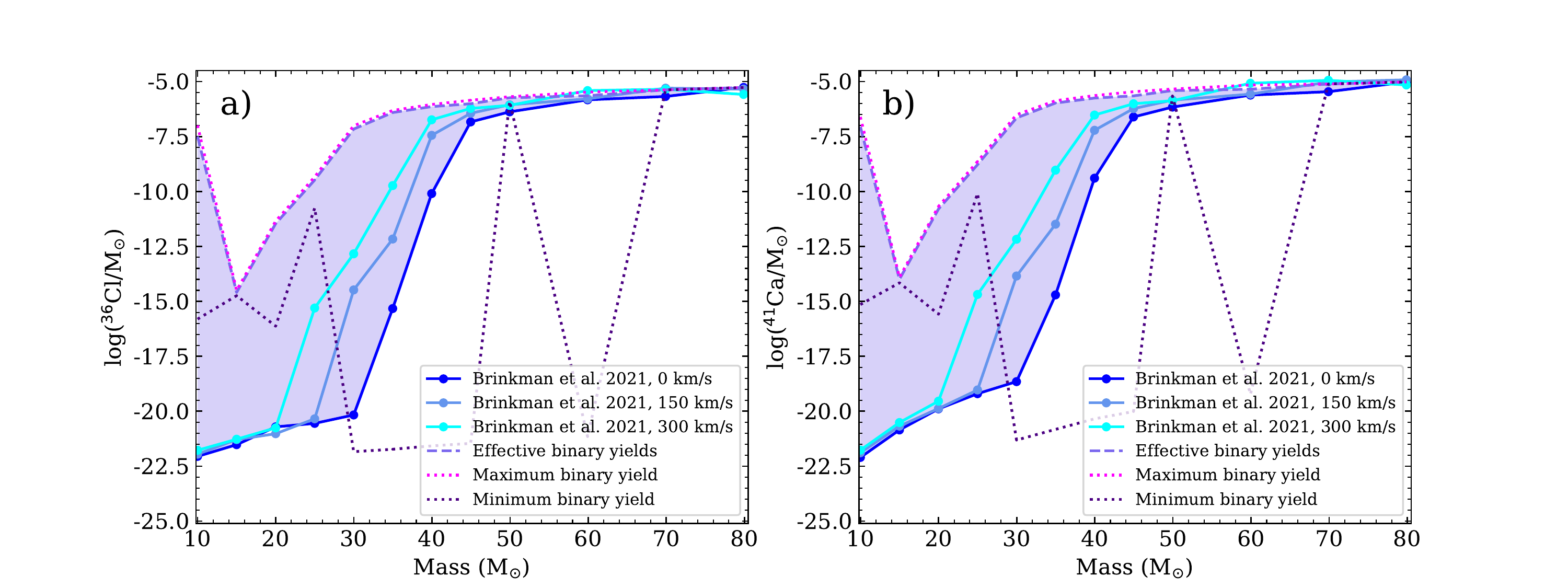}\\
    \caption{The yields of Paper II (blue lines with dots) with the effective binary yield (dashed line), maximum binary yield (purple dotted line), and minimum binary yield (teal dotted line) for the models presented in this work for \Cl{}, and \Ca{} in panels a, and b, respectively. The yields for the individual binary systems are given in Table \ref{BinYields}. The shaded area indicates the potential binary yields, assuming that they are between the effective binary yield and the single non-rotating star yield, as fully non-conservative mass-transfer gives an upper limit to the yields.}
    \label{Set4ClCaFe}
\end{figure*}

\begin{figure*}
    \centering
    \includegraphics[width=0.49\textwidth]{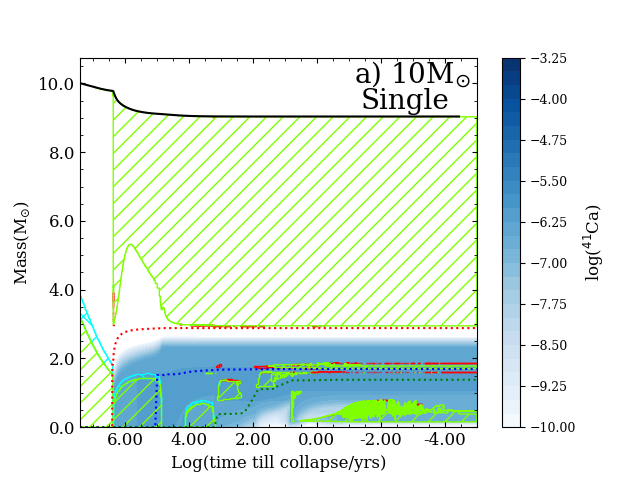}
    \includegraphics[width=0.49\textwidth]{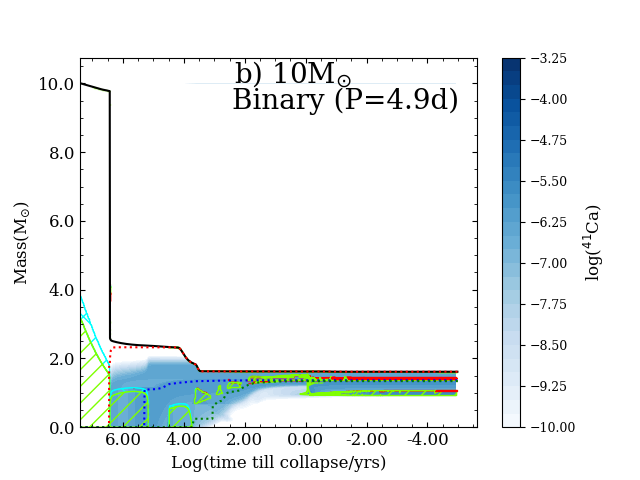}\\
    \includegraphics[width=0.49\textwidth]{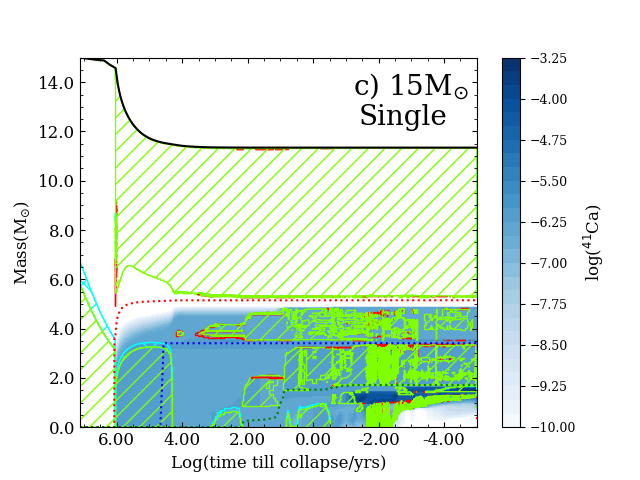}
    \includegraphics[width=0.49\textwidth]{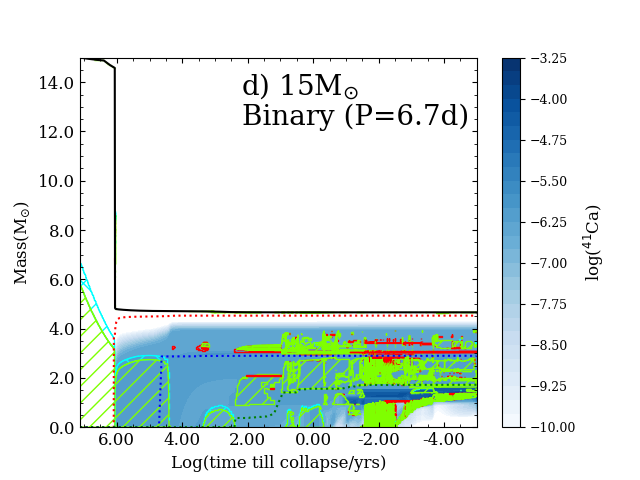}\\
    \includegraphics[width=0.49\textwidth]{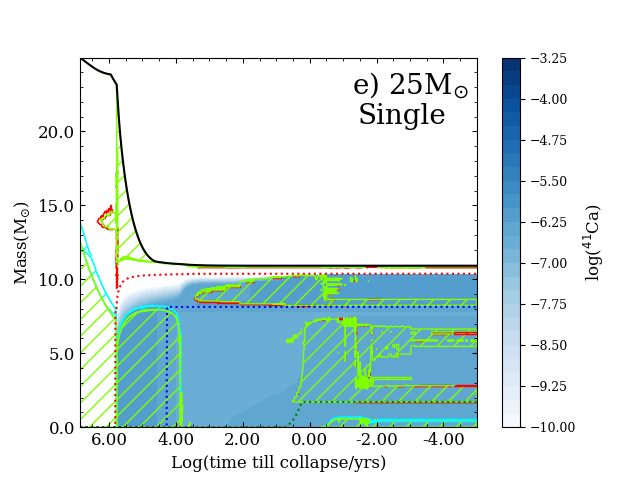}
    \includegraphics[width=0.49\textwidth]{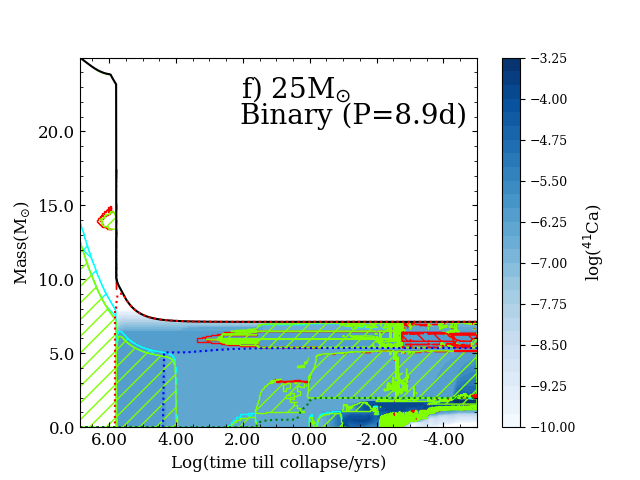}
    \caption{Kippenhahn diagrams for 10, 15, and 25 \msun{}, single stars on the left (panels a, c, and e, respectively), and as primaries of a binary system undergoing Case B mass transfer on the right. The initial periods are 4.9 days for the 10 \msun{} model (panel b), 6.7 days for the 15 \msun{} model (panel d), and 8.9 days for the 25 \msun{} model (panel f). The colour scale shows the \Ca{} mass fraction in the stars. The green shaded areas correspond to areas of convection, the cyan shaded areas to overshooting, and the red shaded areas to semi-convection. The red dotted line indicates the hydrogen depleted core, or helium core, where the hydrogen content is below 0.01 and the helium content is above 0.1. The colour scale shows the \Ca{} mass fraction as a function of the mass coordinate and time.}
    \label{Whyisthereadipat15}
\end{figure*}

Figure \ref{Set4ClCaFe} shows the yields for \Cl{} and \Ca{}, panel a and panel b, respectively. As described in Paper II, a higher initial rotational velocity decreases the initial mass for which the stars become Wolf-Rayet stars, leading to an earlier increase in the yields of these two SLRs. The difference between the maximum/effective binary yield and the minimum binary yield is much larger for these two isotopes than for \Al{}. As mentioned in the previous section, the sharp decrease in the minimum binary yield at 30 \msun{} is caused by the systems with the shortest initial period undergoing a common-envelope phase. As in the case of \Al{}, the binary interactions have the most prominent effect at the lower end of the mass-range discussed in this work. Above $\sim$45 \msun{}, the impact of the binary interactions on the yields decreases, and at $\sim$60 \msun{}, the difference in yields becomes negligible.
For the rotating stars, the effective binary yields become similar to the single star yields at lower initial masses, $\sim$40 \msun{}, as expected.\\
\indent Just as for \Al{}, the effective binary yields for \Cl{} and \Ca{} follow a similar general trend as the single star yields. However, unlike for \Al{}, \Cl{} and \Ca{} experience a strong increase in the yield at 10 \msun{}. The increase at 10 \msun{} is caused by the fact that a deeper layer of the star is reached by the increased mass loss due to the binary interactions. Figure \ref{Whyisthereadipat15} shows the Kippenhahn diagrams (KHDs) for stars with initial masses 10, 15, and 25 \msun{}, single stars on the left, and in a binary system undergoing Case B mass transfer on the right, with the \Ca{} mass fraction on the colour scale (the \Cl{} mass fraction looks similar). The binary system with an initial mass of 10 \msun{} loses more mass than the single star due to binary interactions (see also Table \ref{StellarInfo}). This exposes deeper layers of the star to which \Ca{} (and \Cl{}) has been mixed, as it is produced in the convective He-core. Especially during the final mass loss phase during carbon shell burning (between log(time till collapse/yrs)= 4 and 2), the top layer enhanced in \Ca{} (and \Cl{}) is lost from the primary star. This is due to the fact that these stars experience an additional phase of mass transfer after core helium burning, which results from a strong increase in the stellar radius during this phase. This is a common feature of exposed helium cores with masses of about 2.5 \msun{}, which does not occur for higher masses \citep[see, e.g.,][]{Habets1986Hestars}. This leads to a strong increase in the yield for \Ca{} (and \Cl{}) compared to the single star and a stronger dependence of the yields on the initial period than for the other systems, as can also be seen in the yields of the widest 10 \msun{} binary at a period of 104.6 days. This system does not go through the additional mass-transfer phase and has a significantly lower yield for \Ca{} and \Cl{} than the shorter period systems, though still strongly increased compared to the single star.\\
\indent For the 15 \msun{} star (panels c and d of Figure \ref{Whyisthereadipat15}) the mass loss after the main mass-transfer phase is not strong enough to remove the upper layers of the helium core and the \Ca{} (and \Cl{}) produced in the inner layers is not reached and this star does not experience the additional mass-transfer phase. This leads to smaller yields and thus a smaller increase in the effective binary yield as compared to the 10 \msun{} case, following the general trend of the single stars. For the other single stars, 20-30 \msun{}, the helium burning core is barely reached, as shown by the example of a 25 \msun{} star, and these stars also do not go through an additional mass-transfer phase. The resulting yields are below 10$^{-20}$ \msun{} (see also Table \ref{BinYields}). For these masses, the increased mass loss leads to the exposure of the top of the helium burning core, just as for the 10 \msun{} star, giving a larger effective binary yield, which follows the same trend as the single star yields. Around $\sim$40 \msun{}, the effect of the binary interactions becomes smaller again, especially compared to the models of Paper II including the effects of rotational mixing.
\subsection{Stable isotopes; fluorine-19 and neon-22}
\begin{figure*}
    \centering
\includegraphics[width=\textwidth]{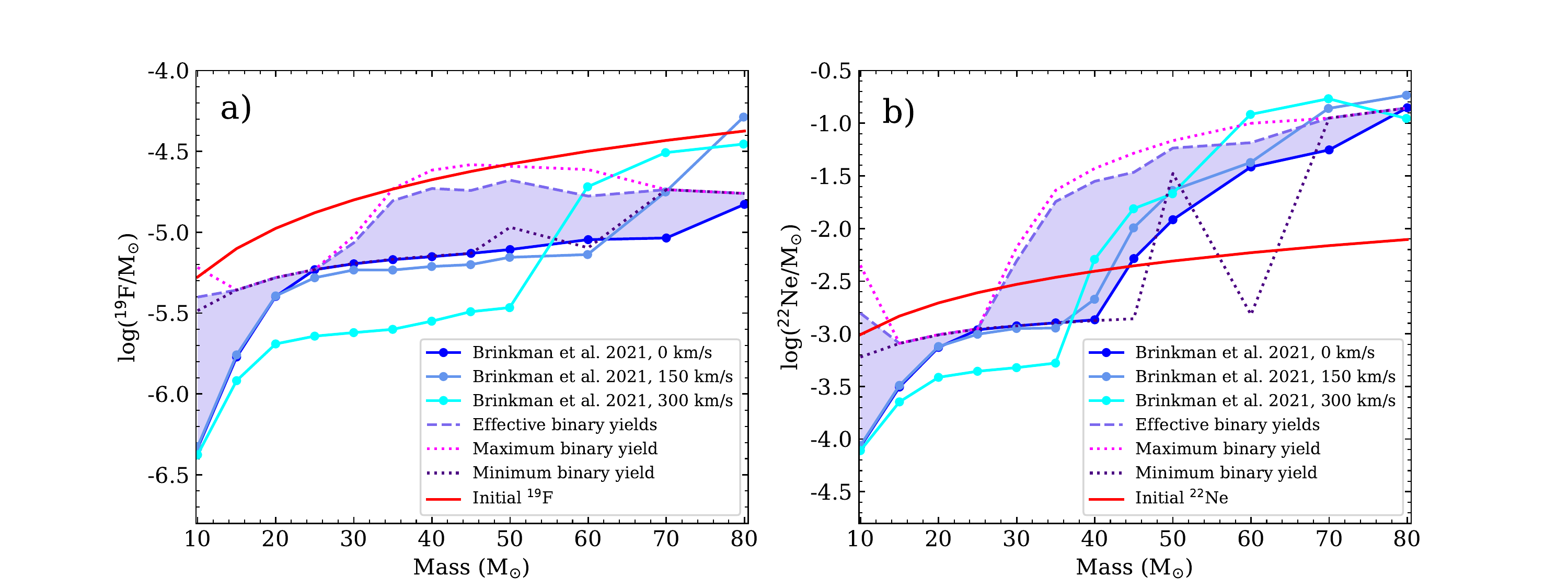}\\
    \caption{The yields for the single star models, both rotating and non-rotating, of Paper II (blue lines with dots) with the effective binary yield (dashed line), maximum binary yield (magenta dotted line), and minimum binary yield (purple dotted line) for the models presented here for \F{} and \Ne{} in panel a, and b, respectively. The yields for the separate binary systems are given in Table \ref{BinYields}. The coloured area indicates the potential binary yields, assuming that they are between the effective binary yield and the single star yield. The red lines in both panels indicate the initial amounts of \F{} and \Ne{} present in the stars.}
    \label{Set4FNe}
\end{figure*}
\begin{figure*}
    \centering
    \includegraphics[width=0.49\textwidth]{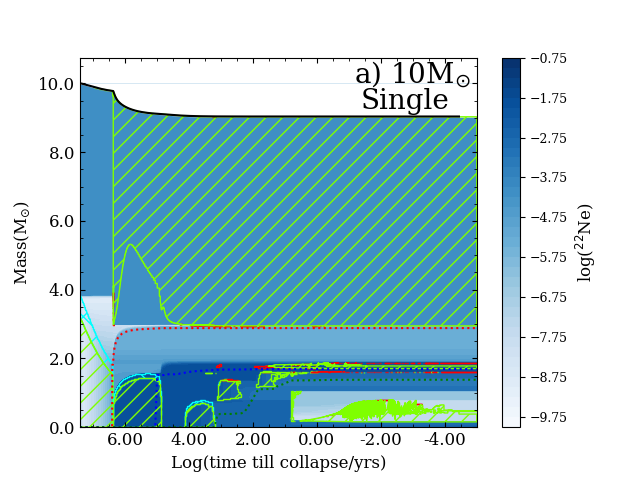}
    \includegraphics[width=0.49\textwidth]{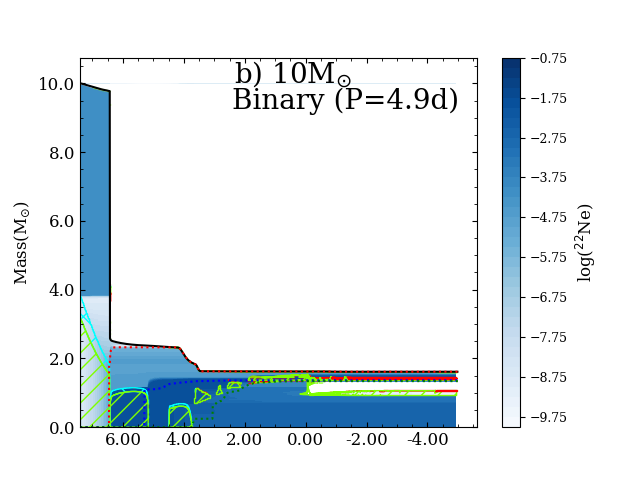}\\
    \includegraphics[width=0.49\textwidth]{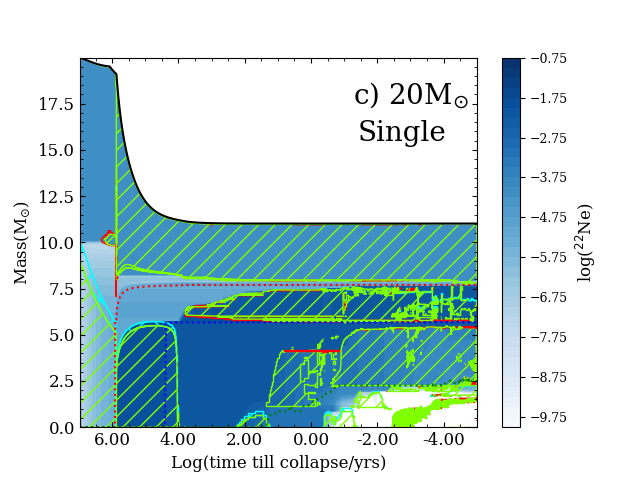}
    \includegraphics[width=0.49\textwidth]{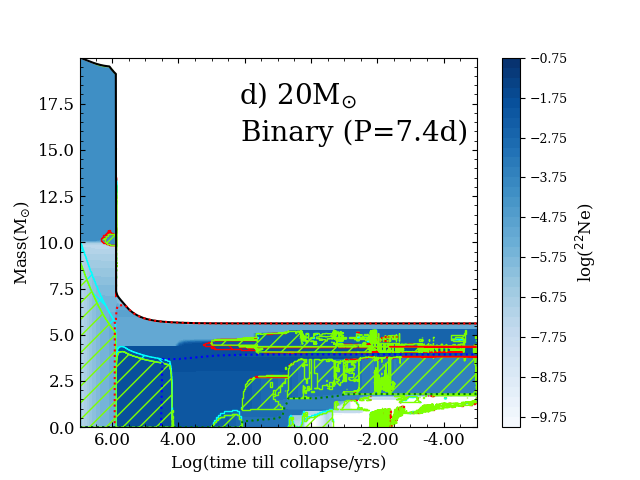}\\
    \includegraphics[width=0.49\textwidth]{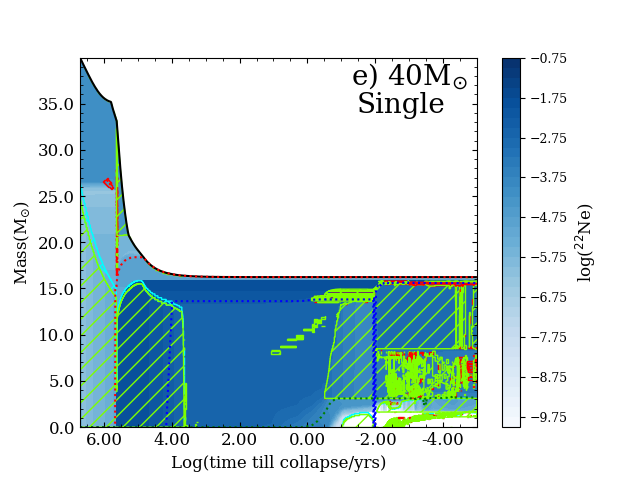}
    \includegraphics[width=0.49\textwidth]{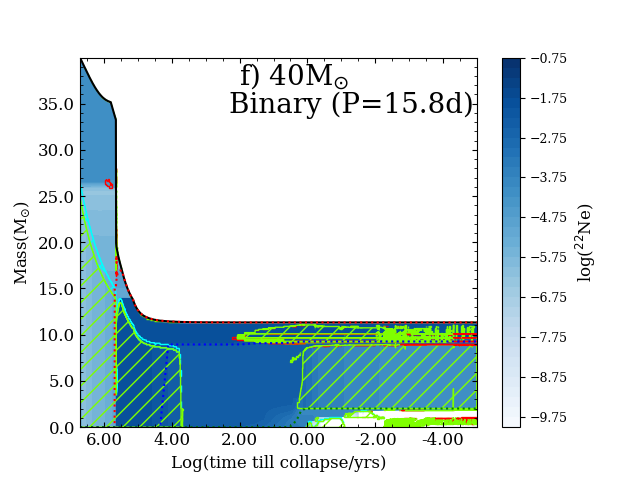}
    \caption{Kippenhahn diagrams for a 10, 20, and 40 \msun{}, single stars on the left (panels a, c, and e), binary stars on the right (panels b, d, and f). The initial periods are 4.9 days for the 10 \msun{} model (panel b), 7.4 days for the 20 \msun{} model (panel d), and 15.8 days for the 40 \msun{} model. All three systems undergo Case B mass-transfer. The colour scale shows the \Ne{}{} mass fraction in the stars. All other colours and shadings are the same as in Figure \ref{Whyisthereadipat15}.}
    \label{Ne22At25}
\end{figure*}

The results for \F{} and \Ne{} are shown in Figure \ref{Set4FNe}. The red line gives the initial mass of the stable isotopes present in the stars at the beginning of the evolution\footnote{In Paper II, the initial masses of the stable isotopes were mistakenly taken from a model that already had slight processing through the CNO-cycle, resulting in a lower mass of \F{} and \Ne{} than based on the metallicity. This has been corrected here.}.\\
For both \F{} and \Ne{}, rotation decreases the yields of the single stars as compared to the non-rotating case, up to 50 \msun{} for the former, and up to 40 \msun{} for the latter, as described in more detail in Paper II.\\
The effective binary yields for both isotopes do not follow the single star trend as clearly as for the SLRs discussed previously. For both isotopes, the effective binary yields are almost identical to the non-rotating single star yields around 25 \msun{}. This is because for both the single star and the binary, all \F{} and \Ne{} present in the envelope is stripped off, but the binary mass loss is not strong enough to reach the helium burning core or the helium burning shell, where these isotopes are produced. For the models with lower initial masses, the added mass loss through the binary interactions does increase the \F{} and \Ne{} yields, and thus the effective binary yields are higher than the single star yields. For \F{} even the increased effective binary yields are still below the initial amount of \F{} present in the star. For \Ne{} the effective binary yield at 10 \msun{} is positive. The increase of the effective binary yields at 10 \msun{} is due to the deeper layers being reached, as seen for \Cl{} and \Ca{} previously. For the models with initial masses above 25 \msun{}, deeper layers of the star are reached, leading to an increase in the effective binary yields again as compared to the single star yields.\\
\indent This is illustrated in Figure \ref{Ne22At25}, showing the KHDs for three initial masses, 10 \msun{}, 20 \msun{}, and 40 \msun{}, single stars on the left, and the primary stars of selected binary systems on the right with the \Ne{} mass fraction on the colour-scale. As already described above for Figure \ref{Whyisthereadipat15}, for the 10 \msun{} star, the primary star loses significantly more mass than the single star, which leads to a large increase of the yield, though either yield is too low to give a positive net yield\footnote{Unlike \Al{}, which is not initially present in the star, there are two types of yields to consider for the stable isotopes, the ``total" yield and the ``net" yield. The total yield is calculated as described above, which ignores the initial amount of the stable isotope present in the star. The net yield is the total yield minus the initial amount of the isotope that was present in the star.}. Only the Case A system at 10 \msun{} gives a positive net yield for \F{}. For the 20 \msun{} models, even though the primary star loses more mass than a single star, the layers uncovered by the binary mass-transfer do not have a large \Ne{} mass fraction and thus the increase between the single star and the binary star is much smaller than for the 10 \msun{} model. Finally, for the 40 \msun{} models, the extra mass loss due to the binary interactions uncovers the deeper layers of the star and the winds strip off the upper layer of the region that belonged to the helium burning core. This increases the yield of the primary star significantly as compared to the single star, leading to a positive net yield for the primary star. By contract, only the fastest rotating single star model has a marginally positive net yield. This shows that while for the SLRs, the impact of the binary interactions already tapers off around 40 \msun{}, for the stable isotopes, the effect is still significant for the higher masses considered here (40-50 \msun{}).\\
\indent For \F{}, only the maximum binary yields of the 35-50 \msun{} models are larger than the initial amount of \F{} present in the star, giving a positive net yield. This is at a significantly lower mass than for the single stars of Paper II, where the only model to give positive \F{} yield is the 80 \msun{} model rotating at 150 km/s. However, only the maximum yields are slightly above the initial amount of \F{} in these stars, which makes massive binaries unlikely candidates to explain the \F{} abundance in the Galaxy.\\
\indent For \Ne{}, only non-rotating single-star models with masses $>$ 45 \msun{} have positive yields, while this minimum mass is $~$ 40 \msun{} for the highest rotation rate. For binary models however, the possible mass range for positive yields is much wider including all stars in our grid with masses greater than 30 \msun{}. Binary interactions may lead to a noticeable increase in the total yields from a given stellar population. To determine the yield increase of a population consisting of binaries versus a population of single stars, we consider a simple test using a Salpeter initial mass function \citep{Salpeter1955IMF}, in which the number of stars of a certain mass is given by:
\begin{equation}
    \frac{dN}{dM} = k \times M^{\alpha}
\end{equation}
where $k$ is a constant determined by the local stellar density, $M$ is the mass of the star in \msun{}, and $\alpha$=-2.35. The total yield of a population can then be expressed as:
\begin{equation}
    Y_{tot} = \int Y(M) \frac{dN}{dM} dM
\end{equation}
where $Y(M)$ is a a function describing how the yield depends on stellar mass. However, because we have only calculated yields at a discrete set of mass values, we replace the integral by a sum over mass bins:
\begin{equation}
    Y_{tot} = \sum _{i} Y_{M,i} N_{M,i}
\end{equation}
where $Y_{M,i}$ is the computed yield in bin $i$ for mass $M$, and $N_{M,i}$ is the number of stars in this bin, given by:
\begin{equation}
    N_{M} = \int_{M_{low,i}}^{M^{up,i}} \frac{dN}{dM}dM = k^{'}(M_{low,i}^{-1.35} - M_{up,i}^{-1.35})  
\end{equation}
 where $k^{'}$ is a constant and M$_{low}$ and M$_{up}$ are the chosen boundaries of a mass-bin based on our sample. We follow this procedure both for a population of single stars (taking for $Y_{M,i}$ the single-star yields) as well as for a population of binary stars, under the assumption that each star is the primary of a binary system and taking for $Y_{M,i}$ the effective binary yields. To get the yield increase of a binary population compared to the population of single stars, we divide the population yields, giving an increase by a factor of 3.95. However, to fully understand the impact of the binary population, a more detailed calculation using galactic chemical evolution models needs to be done (see Section 4.5 of Paper I), which is beyond the scope of this work. Instead, we decided to use a simpler approach that is consistent with our previous works.
\subsection{Helium star winds}
\label{Hestar}
The nucleosynthetic yields from the primary stars of binary systems result from a combination of binary mass-transfer and stellar winds. The stellar winds of massive stars are crucial for the evolution of these stars, but they are also  very uncertain \citep{Smith2014Review}. The final state of the star can be strongly impacted by the choice of mass loss prescription \citep[see, e.g.,][]{Renzo2017}. As noted by \cite{Vink2017}, the mass-loss rates for helium stars (stars without a hydrogen-rich envelope) that are the result of a binary interaction and fall below the limit for Wolf-Rayet stars (5-20 \msun{} for the helium star), are likely not the same as those for actual Wolf-Rayet stars, which are stars that have lost their envelope through their strong stellar winds. The difference in the wind strength is about an order of magnitude \citep[see, e.g.,][Appendix D]{Laplace2021}, and this could potentially impact the nucleosynthetic yields of these stars. To test the impact of this different wind prescription for these stars, we have run models where we changed the wind prescription based on the size of the helium core of the stars. The only systems in our set that are strongly impacted by the change in the mass-loss rates for helium stars are the 10 and 15 \msun{} models, of which especially the 15 \msun{} models might be impacted since their helium cores are around 4-5 \msun{}.\\
\indent To see the impact of the mass-loss rates compared to the models earlier described in the paper (Set 1), we ran two additional sets of models using the reduced mass-loss rates following \cite{Vink2017}. The first set (Set 2) uses the reduced mass-loss rate when the helium core is smaller than 4 \msun{} and is used for the primary stars with 10 and 15 \msun{}. The second set (Set 3) uses the reduced mass-loss rate when the helium core is smaller than 5 \msun{} and is only used for the 15 \msun{} primary stars. This is because the 15 \msun{} models have core masses between 4-5 \msun{}, while those for the 10 \msun{} models are always below 4 \msun{}.\\
\indent Table \ref{HestarYields} gives the yields for the three SLRs for the different initial periods for the 10 and 15 \msun{} models and for the different wind prescriptions. The SLRs and their ratios are impacted by the change in the stellar winds. The stable isotopes instead, \F{} and \Ne{}, are not affected by the changes in the wind. This is because \F{} and \Ne{} are produced in deeper layers than those reached by the stellar winds, for these particular models.  \\
\indent For the 15 \msun{} models, the effects of changing the wind prescription to a reduced wind are clear. For the shortest period binaries (P = 3.8 days), the helium core shrinks below the 4 \msun{} limit, and both Set 2 and Set 3 have the same yields, which are smaller than the yields of Set 1. For the binaries with an initial period of 6.7 and 16.8 days, the yields for Set 2 are comparable to the yields of Set 1, while the yields of Set 3 are smaller. This is because for the wider systems in Set 2 the the Wolf-Rayet wind following \cite{NugisLamers2000} is used, as it is for Set 1, while for the wider systems in Set 3 still the reduced winds are used. The effect of the reduced winds is stronger on \Cl{} and \Ca{} than on \Al{}, which is due to the later production of \Cl{} and \Ca{} and their location deeper under the surface. The stronger the winds, the more likely these layers are reached.\\
\indent For the 10 \msun{} models, the behaviour is less intuitive. While the \Cl{} and \Ca{} yields decrease for the reduced mass loss rate, especially for the widest period, as for Set 3 of the 15 \msun{} models, the \Al{} yields increase for the less efficient winds. This is due to a slight change in the mass-loss history, when the \Al{}-rich layers are close to the surface. For the two widest periods, the `change is minor and the final yields are still close to those of Set 1. For the closest period (P = 2.8 days), the mass-loss increases earlier than for the same model in Set 1, which, combined with the still decaying \Al{} content of the envelope, leads to a strong increase in the \Al{} yield of Set 2 as compared to Set 1.
\section{Early Solar System}\label{sec:Discussion}
The radioactive isotopes described in the previous section were inferred to be present in the early Solar System (ESS) from observed excesses of their daughter nuclei in meteoritic inclusions. To determine whether the binary systems presented in this work can explain the presence of \Al{}, \Cl{}, and \Ca{} in the ESS, we use the simple dilution model for \Al{}, \Cl{}, and \Ca{}, described in Paper II.
\subsection{Selection of the binary systems}
\begin{figure}
    \centering
    \includegraphics[width=0.49\textwidth]{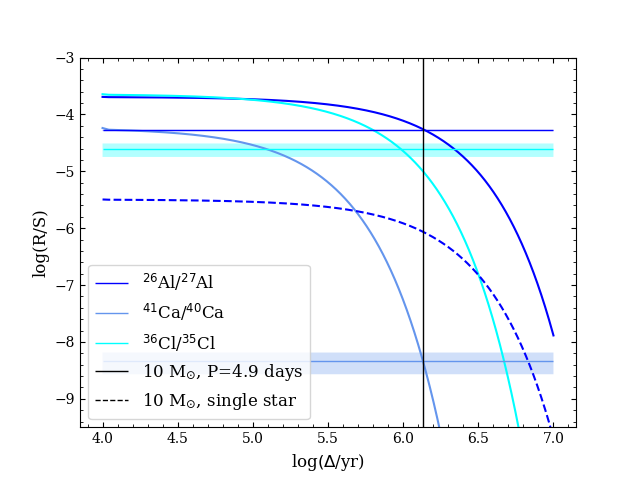}
    \caption{Abundance ratios (R/S) for the three SLRs (R) over their stable reference isotope (S) for a certain dilution factor $f$. The solid lines represent the 10 \msun{} primary star of the system with a period of 4.9 days, the dashed line is the \Al{}/$^{27}$Al-ratio for the 10 \msun{} single star of Paper II. The horizontal bands represent the ESS ratios, with their respective errors. The vertical line represent the delay time for the \Ca{}/$^{40}$Ca-ratio.}
    \label{ESSfigure}
\end{figure}

We apply the same method as in described in Section 5 of Paper II to determine which of the binary systems presented in this paper might be able to explain the abundances of \Al{}, \Cl{}, and \Ca{} in the ESS. Here we repeat the method briefly.\\
\indent We determine a ``dilution factor", $f_{26}$, based on \Al{}. This is defined as $f_{\rm 26}=\frac{M^{\rm ESS}_{\rm 26}}{M^{*}_{\rm 26}}$, where $M^{\rm ESS}_{\rm 26}$ is the mass of \Al{} in the ESS, and $M^{*}_{\rm 26}$ is the mass of \Al{} ejected by the stellar wind, i.e., the total yield. The initial amount of \Al{}, 3.1$\times$10$^{-9}$ \msun{}, is derived by assuming the solar abundance for $^{27}$Al \citep{Lodders2003} and a total mass of 1 \msun{} to be polluted \citep[see][for more details]{Lugaro2018}.\\
\indent This dilution factor for \Al{}, $f_{26}$, is then used to obtain the diluted amount of \Ca{}, which is used to calculate the ``delay time" ($\Delta$t). The delay time can be interpreted as the time interval between wind ejection and the incorporation of the SLRs into the first solids to form the ESS. With the delay time, we reverse decay the initial amount of \Al{} in the ESS, and recalculate $f_{26}$ using the new \Al{} value. We then determine a new amount of \Ca{}. We continue this iteration until we converge to a $\Delta$t within a 10\% difference from the previous value, which gives us a different value of $f_{26}$ for each stellar model. Lastly, we apply the final $f_{26}$ to calculate the diluted amount of \Cl{} and a delay time for \Cl{} as well.\\
\indent With this method, we determined in Paper II that stars above $\sim$40\msun{} can match the ESS \Al{}/$^{27}$Al- and \Ca{}/$^{40}$Ca-ratio, but only the most massive models can also match the \Cl{}/$^{35}$Cl-ratio. For the binaries we see a similar scenario: when considering only \Al{} and \Ca{}, almost all models can match the ESS, with the exception of the models with initial primary masses of 15 and 20 \msun{}, the systems with the shortest periods in the range 25-45 \msun{}, and the widest binary system with an initial mass of 10 \msun{}. To also match \Cl{}, higher initial masses for the primary are needed. The binaries with a primary with an initial mass of 35-45 \msun{} can match all three SLRs for all periods, except the shortest (and the 7.8 days period for the 45 \msun{}). The binaries with initial primary masses of 50 \msun{} and higher can match all three SLRs for all periods. Interestingly, we also find that the binary with an initial primary mass of 10 \msun{} can match the three SLRs for all periods, except the widest of 104.6 days. The 10 \msun{} systems are interesting because the shortest period of this configuration yields a white dwarf, which means that no further pollution is expected from the primary star of this system since the star will not explode and not eject other SLRs, such as \Fe{} and $^{53}$Mn. In Figure \ref{ESSfigure} we show the binary model with an initial primary mass of 10 \msun{} with a period of 4.9 days compared to the isotopic ratios in the ESS. For reference, we also plotted the \Al{}/$^{27}$Al-ratio for the 10 \msun{} single star of Paper II. With a dilution factor of 0.071, both the \Al{}/$^{27}$Al- and \Ca{}/$^{40}$Ca-ratio can be matched with this binary model. The \Cl{}/$^{35}$Cl-ratio is not matched, but is within the range of uncertainties. For the single star model, it is not possible to match any of the ratios. However, the very large dilution factor would imply that the stellar winds comprised 7\% of the total Solar System material, which may be considered as unrealistic. Furthermore, the timescale of the pollution by such a system can prove to be problematic for contributing to the SLR abundances in the ESS. This is because the life-time of these stars, 25-28 Myr, is longer than the lifetime of a Giant Molecular Cloud \citep[see, e.g.,][]{Hartmann2001}, and also because it is at the upper limit of the isolation time for the ESS, as found by \cite{Trueman2022}.
\subsection{Revision of the \Ca{}/$^{40}$Ca-ratio}
\cite{Ku2022Ca} have published a new value for the ESS of \Ca{}/$^{40}$Ca ratio of (2.00$\pm$0.52)$\times$10$^{-8}$, $\sim$4 times higher than the value found by \cite{Liu2017} of (4.6$\pm$1.9)$\times$10$^{-9}$, that was used to determine the results above. When we apply this revised ratio determination to both the models from Paper II and to the models presented here, we find that the new value barely changes which models match the ESS. The main difference is that the delay time is about a factor 20-30\% shorter for the models calculated with the new \Ca{}/$^{40}$Ca value as compared to the older value. This is because matching a higher ratio requires a shorter decay time. Changing the \Ca{}/$^{40}$Ca-ratio does not affect which single stars can match both \Al{} and \Ca{}. For all three SLRs however, the 50 \msun{} model with an initial rotational velocity of 300 km/s is no longer a match. For the binaries, changing the \Ca{}/$^{40}$Ca-ratio does affect which models match all three SLRs, but the models that only match \Al{} and \Ca{}, the number of matches by two - the 25 \msun{} binaries at 6.7 and 71.3 days are no longer a match.
\subsection{Oxygen ratios and C/O wind composition}
There are two more considerations that may be of interest as potential constraints for this scenario of the origin of the three SLRs in the ESS. The first is if the SLR pollution also affects the oxygen isotopic ratios in the ESS, and the second is the carbon-oxygen ratio in the wind. The first is of interest because some CAIs (the FUN class) and some corundum grains are poor in \Al{}, but they have virtually the same oxygen isotopic composition as those that are rich in \Al{} \citep[see, e.g.,][]{Makide2011}. For example, the errors on the measurement of $\Delta$$^{17}$O are of the order of a percent (see Figure 2 of \cite{Makide2011}). Because the oxygen ratios ($^{17}$O/$^{16}$O and $^{18}$O/$^{16}$O) are so well constrained, the model that reproduces the SLRs cannot impact these ratios by more than their respective error-bars, to avoid a correlation between \Al{} and a modification of the oxygen isotopic ratios \citep{GounelleMeibom2007}. 
The composition of the oxygen-isotopes in the winds is dominated by hydrogen and helium burning and its main features are production of $^{16}$O and depletion of $^{17}$O and $^{18}$O, relative to the initial amounts of these isotopes in the star. For the single stars, the $^{17}$O/$^{16}$O and $^{18}$O/$^{16}$O ratios are roughly 8 and 20 times lower than the solar value. Therefore, when we add the oxygen isotopic wind yields of the selected models diluted by $f_{\rm 26}$ to the inferred amount of these isotopes in the Solar System we obtain a decrease of the order of 0.1 to 1.5\% in both the $^{17}$O/$^{16}$O and $^{18}$O/$^{16}$O ratios. When considering the binary models, the $^{17}$O/$^{16}$O and $^{18}$O/$^{16}$O ratios are roughly 13 and 20 times lower than the solar value. As for the single stars, we add the diluted yields of the oxygen isotopes to the solar values, and we find that the ratios decrease in the order of 0.1 to 1.5\% in both the $^{17}$O/$^{16}$O and $^{18}$O/$^{16}$O ratios, as before, and again, with increasing the stellar mass from 30-35 \msun{} to 80 \msun{}.\\
\indent We also checked the C/O-composition of the winds at the time of the ejection of \Cl{} and \Ca{}. This is because the C/O-ratio is an indicator for possible dust formation in the stellar winds. \cite{Dwarkadas2017} states that dust may be needed to incorporate SLRs into the ESS. However, dust formation is only observed in C-rich binary cases \citep[see, e.g.,][]{Lau2020Dust}. Thus the binary models are more relevant in relation to dust formation than single stars. All the binary models that can match all three SLRs have C/O-ratios in their winds of larger than 1 at the end of their evolution, and closer to 1 when the \Al{} yield reaches a value above 10$^{-15}$ \msun{} which is when the isotopes are ejected. The exception are the 50 \msun{} models, where the C/O-ratio is slightly below 1. It depends on how the different elements are incorporated in dust how much the C/O-ratio plays a role in determining the potential source of the SLRs. This is however out of the scope of this work.
\section{Conclusions}\label{sec:Conclusion}
We have investigated the production of the stable isotopes, \F{} and \Ne{}, and radioactive isotopes,\Al{}, \Cl{}, and \Ca{}, in the winds of binary systems undergoing Case A or Case B mass transfer. Then we determined which models could self-consistently explain the ESS abundances of \Al{}, \Cl{}, and \Ca{}. We have found that:
\begin{itemize}
    \item In terms of the structural evolution, the main effect of the binary interactions is an increased mass loss, mostly at the lower end of the mass range investigated in this work. The interactions have a secondary effect on the duration of the burning phases, and also on the core structure at the moment of core collapse. For the lowest mass, the binary interactions can completely change the outcome of the evolution. In fact, our 10 \msun{} binary star ends as a white dwarf or an electron-capture supernova instead of as an iron core-collapse supernova.
    \item For the short-lived radioactive isotopes, it is mostly the WRs in the mass-range 40-80 \msun{} that give significant yields. For \Al{}, the binary interactions lose their impact above $\sim$50 \msun{}, while for \Cl{} and \Ca{}, this happens at slightly higher masses, $\sim$60 \msun{}.
    \item Only a very narrow range of initial primary masses and periods produce a net positive \F{} yield (the maximum yields for the 35-50 \msun{} models).
    \item For \Ne{} most systems with initial masses above 30 \msun{}, except for the shortest periods in the range 35-45 \msun{}, produce positive net yields. With these systems we found that a population of binaries can produce about a factor of 4 times more \Ne{} as compared to a population of only single stars considering a Salpeter IMF.
    \item In Section \ref{sec:Discussion}, we have investigated which of the stellar models described in this paper could explain the ESS abundances. Depending on the initial primary mass and initial period, only stars with an initial mass of 15 and 20 \msun{} do not explain the \Al{} and \Ca{} abundances, along with the shortest periods for 25-45 \msun{}, and the widest system at 10 \msun{}. All the models with mass $\geq$50 \msun{} can also explain the \Cl{} abundances. The 10 \msun{} models that do not become iron core-collapse supernovae, but white dwarfs or electron-capture supernovae have very different final yields, which needs to be further investigated, also in relation to the SLRs.
\end{itemize}
\indent A more detailed analysis of several uncertainties should be performed, which includes different prescriptions for the winds and for the rotational boost on wind loss, as well as investigations of the effect of reaction rate uncertainties specifically on the destruction of \F{}, \Ca{}, and \Cl{}, and the neutron source $^{22}$Ne($\alpha$,n)$^{25}$Mg reaction \citep{2021PhRvCAdsley}. The interaction between rotation and binarity  will need further investigation, but it is difficult to preform due to the complexity of the angular momentum coupling between the two stars. Also, a more detailed analysis of the uncertainties of binary evolution and their effects on the yields should be performed. These uncertainties include, but are not limited to, the mass-transfer efficiency, which includes accretion onto the secondary and potentially on the primary in a case of reverse mass transfer, the formation of common envelopes and their effect on the binary parameters, and variations in the initial mass-ratio between the stars. Finally, to present a complete view, the explosive nucleosynthetic yields will need to be calculated using our models as the progenitors of the explosion. 
\section*{Acknowledgements} \label{sec:thanks}
HEB thanks the MESA team for making their code publicly available.  HEB, MP, and ML acknowledge the support from the ERC Consolidator Grant (Hungary) program (RADIOSTAR, G.A. n. 724560). HEB acknowledges support from the Research Foundation Flanders (FWO) under grant agreement G089422N. MP acknowledge the support to NuGrid from JINA-CEE (NSF Grant PHY-1430152) and STFC (through the University of Hull's Consolidated Grant ST/R000840/1), and ongoing access to {\tt viper}, the University of Hull High Performance Computing Facility. MP and ML acknowledge the "Lend{\"u}let-2014" Program of the Hungarian Academy of Sciences (Hungary) for support. This work was supported by the European Union's Horizon 2020 research and innovation program (ChETEC-INFRA -- Project no. 101008324), and the IReNA network supported by US NSF AccelNet (Grant No. OISE-1927130). 
\software{MESA (\citealt{MESA1, MESA2, MESA3, MESA4})}
\bibliographystyle{apj}
\bibliography{references}
\appendix
\section{}\label{StellarEvolTable}
\begin{table*}[b]
\caption{Selected details of the evolution of the stellar models: M$_{\rm ini}$ is the initial mass in M$_{\odot}$; P$_{\rm ini}$ the initial period in days; t$_{\rm H}$, t$_{\rm He}$, and t$_{\rm tot}$ the duration of hydrogen burning, helium burning, and the total evolution time in Myr, respectively. M$_{*,\rm H}$, M$_{*,\rm He}$, and M$_{*,\rm C}$ are the masses of the stars at the end of their respective burning phases; M$_{c,\rm He}$ and M$_{c,\rm C}$ the masses of the hydrogen-depleted core and the helium-depleted core, respectively, at the end of the corresponding burning phases in M$_{\odot}$; and $\Delta$M the total mass lost in M$_{\odot}$. The single star models were taken from Paper\,II. The periods printed in boldface indicate the models used as representative for Case A and Case B mass-transfer in Section \ref{Results1}. Case indicates the first mass transfer the system undergoes, which does not include further mass-transfer phases such as Case AB or BC.}
\begin{center}\begin{tabular}{ccc|ccc|ccc|ccc}
\hline 
M$ _{\rm ini} $ &  P$ _{\rm ini} $&Case& t$ _{\rm H} $ & M$ _{c, \rm He} $ &M$_{*,\rm H}$&t$ _{\rm He} $ & M$ _{c,\rm C} $&M$_{*,\rm He}$ &  M$_{*,\rm C}$ &t$ _{\rm tot} $ &$  \Delta$M\\
(M$ _{\odot}$) & (days) &-& (Myr)& (M$ _{\odot} $)& (M$ _{\odot} $)& (Myr)& (M$ _{\odot} $)& (M$ _{\odot} $)&(M$ _{\odot} $) & (Myr)& (M$ _{\odot} $)\\
\hline
10	&-$^{1}$                &-	&23	     &1.83	&9.78	&1.98	&1.52	&9.11	&9.04	&25.37	&0.96\\
    &\textbf{2.8}$^{2,5}$  	&A	&24.1	 &1.11	&4.17	&3.44	&1.10	&1.86	&1.85	&28.18	&8.71\\
    &4.9$^{3}$             	&B	&22.98	 &1.83	&9.78	&2.38	&1.26	&2.33	&1.93	&25.73	&8.39\\
    &\textbf{13.1}$^{3}$   	&B	&22.98	 &1.83	&9.78	&2.33	&1.27	&2.37	&2.26	&25.68	&8.18\\
    &104.6$^{3}$           	&B	&22.98	 &1.83	&9.78	&2.25	&1.31	&2.46	&2.44	&25.60	&7.56\\
 \hline											
15	 &-                     &-	&12.15	&3.62	&14.58	&0.96	&3.41	&11.49	&11.35	&13.26	&3.65\\
    &\textbf{3.8}          	&A	&12.40	&2.66	&6.89	&1.22	&1.97	&3.45	&3.43	&13.81	&11.56\\
    &6.7	                &B	&12.15	&3.62	&14.58	&0.99	&2.35	&4.07	&4.01	&13.31	&10.99\\
    &\textbf{16.8}         	&B	&12.15	&3.62	&14.58	&0.99	&2.39	&4.11	&4.05	&13.3	&10.95\\
    \hline
20	    &	 -                     	&	-	&	8.53	&	5.73	&	19.12	&	0.66	&	5.66	&	11.24	&	11.02	&	9.29	&	8.98	\\
    	&	 2.5 $^{1,5}$          	&	A	&	9.11	&	4.15	&	9.74	&	0.76	&	3.14	&	4.95	&	4.95	&	10.10	&	15.05	\\
    	&	 \textbf{5.1}$^{1}$    	&	A	&	8.61	&	4.74	&	9.88	&	0.74	&	3.28	&	5.18	&	5.13	&	9.45	&	14.87	\\
    	&	6.2	&	B	&	8.53	&	5.73	&	19.12	&	0.68	&	3.67	&	5.66	&	5.6	&	9.31	&	14.39	\\
    	&	7.4	&	B	&	8.53	&	5.74	&	19.12	&	0.68	&	3.69	&	5.68	&	5.61	&	9.31	&	14.38	\\
    	&	 \textbf{18.4}         	&	B	&	8.53	&	5.74	&	19.12	&	0.67	&	3.74	&	5.76	&	5.67	&	9.3	&	14.33	\\
    	&	66.2	&	B	&	8.53	&	5.74	&	19.12	&	0.67	&	3.8	&	5.79	&	5.73	&	9.3	&	14.26	\\
    	&	132.4	&	B	&	8.53	&	5.74	&	19.12	&	0.67	&	3.8	&	5.81	&	5.74	&	9.3	&	14.25	\\
  \hline
25	&	 -$^{1}$               	&	-	&	6.8	&	7.99	&	23.2	&	0.53	&	8.13	&	11.04	&	10.92	&	7.41	&	14.07	\\
    	&	 2.7$^{5}$             	&	A	&	7.17	&	6.27	&	13.02	&	0.64	&	4.18	&	6.14	&	6.08	&	7.90	&	18.91	\\
    	&	 \textbf{6.7}$^{5}$    	&	A	&	6.84	&	7.1	&	13.09	&	0.57	&	4.71	&	6.82	&	6.75	&	7.49	&	18.24	\\
    	&	8.9	&	B	&	6.8	&	7.99	&	23.2	&	0.55	&	5.06	&	7.21	&	7.12	&	7.42	&	17.87	\\
    	&	 \textbf{17.8}         	&	B	&	6.8	&	7.99	&	23.2	&	0.56	&	5.09	&	7.24	&	7.18	&	7.42	&	17.83	\\
    	&	 71.3$^{1}$            	&	B	&	6.8	&	7.99	&	23.2	&	0.54	&	5.15	&	7.31	&	7.23	&	7.42	&	17.76	\\
  \hline
30	&	 -$^{1}$               	&	-	&	5.8	&	10.35	&	26.89	&	0.47	&	10.73	&	13.56	&	13.47	&	6.32	&	16.52	\\
    	&	 2.8$^{4}$             	&	A	&	 -    	&	 -    	&	 -    	&	 -   	&	 -    	&	 -    	&	-      	&	 -     	&	-     	\\
    	&	 \textbf{8.4}$^{5}$    	&	A	&	5.82	&	9.55	&	16.59	&	0.48	&	6.5	&	8.55	&	8.41	&	6.36	&	21.57	\\
    	&	 10.1$^{5}$          	&	B	&	5.8	&	10.32	&	26.81	&	0.48	&	6.59	&	8.69	&	8.56	&	6.34	&	21.43	\\
    	&	 12.2$^{5}$            	&	B	&	5.8	&	10.32	&	26.88	&	0.48	&	6.58	&	8.67	&	8.53	&	6.34	&	21.46	\\
    	&	 \textbf{30.3}$^{5}$   	&	B	&	5.8	&	10.32	&	26.88	&	0.47	&	6.64	&	8.93	&	8.68	&	6.33	&	21.31	\\
    	&	 75.4$^{5}$            	&	B	&	5.8	&	10.32	&	26.88	&	0.47	&	6.67	&	8.79	&	8.66	&	6.33	&	21.33	\\
\hline
\end{tabular}\end{center}
\label{StellarInfo}
\end{table*}
\begin{table*}
\centering {Table \ref{StellarInfo} continued.}
\begin{center}\begin{tabular}{ccc|ccc|ccc|ccc}
\hline 
M$ _{\rm ini} $ &  P$ _{\rm ini} $&Case& t$ _{\rm H} $ & M$ _{c, \rm He} $ &M$_{*,\rm H}$&t$ _{\rm He} $ & M$ _{c,\rm C} $&M$_{*,\rm He}$ &  M$_{*,\rm C}$ &t$ _{\rm tot} $ &$  \Delta$M\\
(M$ _{\odot}$) & (days) &-& (Myr)& (M$ _{\odot} $)& (M$ _{\odot} $)& (Myr)& (M$ _{\odot} $)& (M$ _{\odot} $)&(M$ _{\odot} $) & (Myr)& (M$ _{\odot} $)\\
\hline
35	&	 -                     	&	-	&	5.15	&	12.74	&	30.38	&	0.42	&	12.95	&	15.64	&	15.46	&	5.62	&	19.51	\\
    	&	 2.9$^{4}$             	&	A	&	 -    	&	 -    	&	 -    	&	 -   	&	 -    	&	 -    	&	-      	&	 -   	&	-      	\\
    	&	 \textbf{8.8}$^{5}$    	&	A	&	5.17	&	12.02	&	20.73	&	0.44	&	7.32	&	9.64	&	9.57	&	5.66	&	25.4	\\
    	&	 10.6$^{5}$            	&	A	&	5.16	&	12.63	&	21.23	&	0.43	&	7.62	&	9.92	&	9.85	&	5.64	&	25.13	\\
    	&	 12.7$^{5}$            	&	B	&	5.16	&	12.66	&	30.42	&	0.43	&	7.57	&	9.89	&	9.81	&	5.64	&	25.17	\\
    	&	 \textbf{31.5}$^{5}$   	&	B	&	5.16	&	12.66	&	30.42	&	0.43	&	7.81	&	10.14	&	10.08	&	5.64	&	24.9	\\
    	&	 78.6$^{5}$            	&	B	&	5.16	&	12.66	&	30.42	&	0.43	&	8.05	&	10.42	&	10.35	&	5.63	&	24.63	\\
\hline	
40	&	 -                     	&	- 	&	4.69	&	15.17	&	33.23	&	0.39	&	13.63	&	16.4	&	16.24	&	5.12	&	23.74	\\
    	&	 3.1$^{4}$             	&	A 	&	 -   	&	 -    	&	 -    	&	 -   	&	 -    	&	 -    	&	-      	&	 -   	&	-       	\\
    	&	 \textbf{7.6}$^{5}$    	&	A 	&	4.72	&	14.43	&	24.83	&	0.41	&	8.5	&	10.9	&	10.85	&	5.18	&	29.12	\\
    	&	 15.8$^{5}$            	&	B 	&	4.7	&	15.05	&	33.3	&	0.41	&	8.95	&	11.39	&	11.33	&	5.15	&	28.64	\\
    	&	 20.4$^{5}$            	&	B 	&	4.7	&	15.05	&	33.3	&	0.4	&	9.03	&	11.42	&	11.36	&	5.15	&	28.61	\\
    	&	 \textbf{32.8}$^{5}$   	&	B 	&	4.7	&	15.05	&	33.29	&	0.4	&	9.11	&	11.57	&	11.51	&	5.15	&	28.45	\\
    	&	 81.7$^{5}$            	&	B 	&	4.7	&	15.05	&	33.3	&	0.4	&	10.09	&	12.61	&	12.55	&	5.14	&	27.41	\\
    \hline	
45	&	 -                     	&	-  	&	4.35	&	17.57	&	35.74	&	0.36	&	14.85	&	17.95	&	17.9	&	4.76	&	27.06	\\
    	&	 3.2$^{4}$             	&	 A 	&	  -  	&	 -    	&	 -    	&	 -   	&	 -    	&	 -    	&	-     	&	 -   	&	-       	\\
    	&	 \textbf{6.5}$^{5}$    	&	 A 	&	4.38	&	16.85	&	28.6	&	0.39	&	9.67	&	12.26	&	12.18	&	4.81	&	32.77	\\
    	&	 7.8$^{1}$             	&	 A 	&	4.37	&	16.96	&	28.91	&	 -   	&	 -    	&	-     	&	-       	&	4.51	&	-     	\\
    	&	 19.5$^{5}$            	&	 B 	&	4.35	&	17.46	&	35.91	&	0.39	&	10.37	&	12.86	&	12.8	&	4.79	&	32.15	\\
    	&	 23.4$^{5}$            	&	 B 	&	4.35	&	17.46	&	35.95	&	0.38	&	10.38	&	12.92	&	12.84	&	4.79	&	32.11	\\
    	&	 \textbf{42.0}$^{5}$   	&	 B 	&	4.35	&	17.46	&	35.95	&	0.38	&	10.56	&	13.23	&	13.17	&	4.78	&	31.79	\\
    	&	 69.9$^{5}$            	&	 B 	&	4.35	&	17.46	&	35.95	&	0.37	&	11.36	&	14.09	&	14.02	&	4.78	&	30.93	\\
\hline
50	&	 -                     	&	 - 	&	4.09	&	20.05	&	37.93	&	0.35	&	16.84	&	19.92	&	19.87	&	4.48	&	30.07	\\
    	&	 \textbf{8.1}$^{5}$    	&	 A 	&	4.1	&	19.43	&	32.72	&	0.37	&	11.09	&	13.63	&	13.56	&	4.52	&	36.39	\\
    	&	 14.0$^{5}$            	&	 A 	&	4.09	&	19.61	&	33.1	&	0.37	&	11.2	&	13.82	&	13.74	&	4.51	&	36.21	\\
    	&	 21.7$^{5}$            	&	 B 	&	4.09	&	19.77	&	35.3	&	0.36	&	11.43	&	14.09	&	14	&	4.5	&	35.94	\\
    	&	 \textbf{29.1}$^{5}$   	&	 B 	&	4.09	&	19.85	&	38.23	&	0.37	&	11.61	&	14.31	&	14.23	&	4.5	&	35.71	\\
    	&	 72.3$^{5}$            	&	 B 	&	4.09	&	19.85	&	38.29	&	0.36	&	12.83	&	15.69	&	15.61	&	4.49	&	34.33	\\
    	&	 144.6$^{5}$           	&	 B 	&	4.09	&	19.85	&	38.29	&	0.35	&	14.66	&	17.55	&	17.47	&	4.48	&	32.47	\\
    \hline
60	&	 -                     	&	-   	&	3.71	&	25.04	&	41.39	&	0.33	&	19.49	&	22.76	&	22.67	&	4.07	&	37.25	\\
    	&	 3.5$^{4}$             	&	 A  	&	 -   	&	 -    	&	 -    	&	 -   	&	 -    	&	 -    	&	 -     	&	 -   	&	-  	\\
    	&	 \textbf{7.2}$^{5}$    	&	 A  	&	3.71	&	24.53	&	40.22	&	0.33	&	19.24	&	22.53	&	22.43	&	4.07	&	37.49	\\
    	&	 14.9$^{5}$            	&	 B  	&	3.7	&	24.69	&	41.76	&	0.34	&	13.68	&	16.47	&	16.38	&	4.08	&	43.54	\\
    	&	 17.8$^{5}$            	&	 B  	&	3.7	&	24.69	&	41.76	&	0.34	&	13.76	&	16.55	&	16.47	&	4.08	&	43.46	\\
    	&	 \textbf{37.0}$^{5}$   	&	 B  	&	3.7	&	24.69	&	41.76	&	0.34	&	14.09	&	16.98	&	16.9	&	4.08	&	43.02	\\
    	&	 92.2$^{5}$            	&	 B  	&	3.7	&	24.69	&	41.76	&	0.33	&	17.18	&	20.3	&	20.22	&	4.07	&	39.71	\\
\hline
70	&	 -$^{6}$               	&	-  	&	3.43	&	30.29	&	50.32	&	0.31	&	22.86	&	26.36	&	26.25	&	3.78	&	43.65	\\
    	&	 \textbf{39.1}$^{5}$   	&	 B 	&	3.43	&	29.74	&	50.14	&	0.32	&	17.28	&	20.54	&	20.46	&	3.79	&	49.45	\\
\hline
80	&	 -                      	&	-  	&	3.24	&	34.98	&	50.72	&	0.31	&	20.68	&	24.1	&	23.99	&	3.58	&	55.9	\\
   	&	 \textbf{33.9}$^{5}$    	&	 B 	&	3.23	&	34.69	&	54.77	&	0.31	&	19.89	&	23.09	&	23.01	&	3.58	&	56.88	\\

\end{tabular}\end{center}
\raggedright
$^{1}$ This run was terminated before the core collapse due to numerical difficulties.\\
$^{2}$ This primary star has lost such a significant amount of mass that its final state will be a white dwarf. At the end of the simulation the remaining stellar mass is 1.30\msun{}.\\
$^{3}$ The final core mass of this star is such that it is a potential electron-capture supernova.\\
$^{4}$ Terminated due to the formation of a common envelope.\\
$^{5}$ The primary star of this system was uncoupled and further evolved as a single star as the secondary overflows its Roche lobe.\\
$^{6}$ This run experienced computational difficulties in the final phases, leading to a much larger M$_{c,\rm O}$ than for any of the other models.
\end{table*}
\newpage
\section{}\label{YieldsTable}
\begin{table*}[b]
\caption{Wind yields in \msun{} for the binary models for the key isotopes \F{}, \Ne{}, \Al{}, \Cl{}, and \Ca{}. $M_{\rm ini}$ is the initial mass in M$_{\odot}$ and $P_{\rm ini}$ is the initial period of the binary system in days. For the stable isotopes, the initial amount present in the star is given. For the radioactive isotopes, the yields are not corrected for radioactive decay that might take place during the evolution of the star. The top lines for each mass give the non-rotating single star yields from Paper\,II. The periods printed in boldface indicate the models used as representative for Case A and Case B mass-transfer in Section \ref{Results1}.}
\begin{center}
\begin{tabular}{cc|cc|ccc|ccc}
\hline 
$M _{\rm ini} $ &  $P _{\rm ini} $& $^{19}$F$_{\rm ini}$ & $^{19}$F &$^{22}$Ne$_{\rm ini}$ & $^{22}$Ne&$^{26}$Al & $^{36}$Cl& $^{41}$Ca\\
(M$ _{\odot}$)& (days) &(M$ _{\odot}$) &(M$ _{\odot}$) &(M$ _{\odot}$) &(M$ _{\odot}$) &(M$ _{\odot}$) &(M$ _{\odot}$) &(M$ _{\odot}$)\\
\hline
10 & -$^{1}$            &5.28e-06 &4.61e-07 &9.82e-4 &8.47e-05&7.19e-11 &8.82e-23 &7.90e-23\\
 & \textbf{2.8}$^{2,5}$   &5.28e-06 &6.06e-6 &9.82e-4 &3.74e-08 &3.74e-08 &1.01e-07 &2.44e-07\\
 & 4.9$^{3}$            &5.28e-06 &3.27e-06 &9.82e-4 &6.07e-4 &1.02e-07 &2.05e-08 &8.45e-08\\
 & \textbf{13.1}$^{3}$  &5.28e-06 &3.27e-06 &9.82e-4 &6.04e-4 &1.01e-07 &4.61e-09 &2.04e-08\\
 & 104.6$^{3}$          &5.28e-06 &3.28e-06 &9.82e-4 &6.01e-4 &2.16e-08 &1.56e-16 &7.15e-16\\
\hline
15 & -                  &7.93e-06 &1.69e-06 &1.47e-3 &3.12e-4 &8.85e-09 &3.00e-22 &1.41e-21\\
   & \textbf{3.8}       &7.93e-06 &4.39e-06 &1.47e-3 &8.13e-4 &6.04e-07 &1.81e-15 &6.96e-15\\
   & 6.7                &7.93e-06 &4.39e-06 &1.47e-3 &8.11e-4 &4.71e-07 &3.12e-15 &1.24e-14\\
   & \textbf{16.8}      &7.93e-06 &4.39e-06 &1.47e-3 &8.10e-4 &4.58e-07 &2.73e-15 &1.09e-14\\
\hline
20 & -                  &1.06e-05 &4.01e-06 &1.96e-3 &7.43e-4 &1.81e-07 &1.94e-21 &1.28e-20\\
   & 2.5$^{1,5}$        &1.06e-05 &5.24e-06 &1.96e-3 &9.74e-4 &1.53e-06 &7.43e-17 &2.60e-16\\
   & \textbf{5.1}$^{1}$ &1.06e-05 &5.24e-06 &1.96e-3 &9.83e-4 &2.42e-06 &3.05e-12 &1.41e-11\\
   & 6.2                &1.06e-05 &5.24e-06 &1.96e-3 &9.90e-4 &2.13e-06 &4.47e-12 &2.08e-11\\
   & 7.4                &1.06e-05 &5.24e-06 &1.96e-3 &9.80e-4 &2.12e-06 &4.44e-12 &2.07e-11\\
   & \textbf{18.4}      &1.06e-05 &5.24e-06 &1.96e-3 &9.80e-4 &2.07e-06 &4.27e-12 &1.99e-11\\
   & 66.2               &1.06e-05 &5.24e-06 &1.96e-3 &9.81e-4 &2.01e-06 &3.85e-12 &1.79e-11\\
   & 132.4              &1.06e-05 &5.24e-06 &1.96e-3 &9.81e-4 &2.00e-06 &3.84e-12 &1.79e-11\\
 \hline
25 & -$^{1}$            &1.32e-05 &5.87e-06 &2.46e-3 &1.09e-3 &1.17e-06 &2.80e-21 &6.27e-20\\
   & 2.7$^{5}$          &1.32e-05 &5.88e-06 &2.46e-3 &1.11e-3 &5.11e-06 &1.75e-11 &8.10e-11\\
   & \textbf{6.7}$^{5}$ &1.32e-05 &5.88e-06 &2.46e-3 &1.11e-3 &6.31e-06 &3.46e-10 &1.72e-09\\
   & 8.9                &1.32e-05 &5.88e-06 &2.46e-3 &1.11e-3 &5.93e-06 &4.60e-10 &2.30e-09\\
   & \textbf{17.8}      &1.32e-05 &5.88e-06 &2.46e-3 &1.11e-3 &5.86e-06 &4.41e-10 &2.20e-09\\
   & 71.3$^{1}$         &1.32e-05 &5.88e-06 &2.46e-3 &1.11e-3 &5.76e-06 &3.94e-10 &1.96e-09\\
 \hline
30 & -$^{1}$            &1.59e-05 &6.38e-06 &2.95e-3 &1.19e-3 &3.41e-06 &6.76e-21 &2.22e-19\\
   & 2.8$^{4}$          &1.59e-05 &6.41e-06 &2.95e-3 &1.19e-3 &5.59e-07 &1.44e-22 &4.87e-22\\
   & \textbf{8.4}$^{5}$ &1.59e-05 &9.46e-06 &2.95e-3 &6.63e-3 &1.17e-05 &9.78e-08 &3.11e-07\\
   & 10.1$^{5}$       &1.59e-05 &8.98e-06 &2.95e-3 &5.46e-3 &1.17e-05 &7.88e-08 &2.57e-07\\
   & 12.2$^{5}$         &1.59e-05 &9.06e-06 &2.95e-3 &5.53e-3 &1.18e-05 &7.94e-08 &2.60e-07\\
   &\textbf{30.3}$^{5}$ &1.59e-05 &8.42e-06 &2.95e-3 &4.67e-3 &1.16e-05 &6.74e-08 &2.20e-07\\
   & 75.4$^{5}$         &1.59e-05 &9.33e-06 &2.95e-3 &6.05e-3 &1.15e-05 &8.95e-08 &2.87e-07\\
   \hline
\end{tabular}\end{center}
\label{BinYields}
\end{table*}
\begin{table*}[t]
\centering {Table \ref{BinYields} continued.}
\begin{center}\begin{tabular}{cc|cc|cc|ccc}
\hline 
$M _{\rm ini} $ &  $P _{\rm ini} $& $^{19}$F$_{\rm ini}$ & $^{19}$F &$^{22}$Ne$_{\rm ini}$ & $^{22}$Ne&$^{26}$Al & $^{36}$Cl& $^{41}$Ca\\
(M$ _{\odot}$)& (days) &(M$ _{\odot}$) &(M$ _{\odot}$) &(M$ _{\odot}$) &(M$ _{\odot}$) &(M$ _{\odot}$) &(M$ _{\odot}$) &(M$ _{\odot}$)\\
\hline
35 & -                  &1.85e-05 &6.77e-06 &3.44e-3 &1.27e-3 &8.44e-06 &4.70e-16 &1.96e-15\\
   & 2.9$^{4}$          &1.85e-05 &6.80e-06 &3.44e-3 &1.27e-3 &1.20e-06 &1.86e-22 &1.50e-21 \\
   & \textbf{8.8}$^{5}$ &1.85e-05 &1.84e-05 &3.44e-3 &2.31e-2 &1.87e-05 &4.84e-07 &1.35e-06 \\
   & 10.6$^{5}$         &1.85e-05 &1.81e-05 &3.44e-3 &2.24e-2 &1.82e-05 &4.83e-07 &1.31e-06 \\
   & 12.7$^{5}$         &1.85e-05 &1.86e-05 &3.44e-3 &2.21e-2 &1.84e-05 &4.75e-07 &1.30e-06 \\
   & \textbf{31.5}$^{5}$&1.85e-05 &1.68e-05 &3.44e-3 &2.03e-2 &1.79e-05 &4.55e-07 &1.21e-06 \\
   & 78.6$^{5}$         &1.85e-05 &1.54e-05 &3.44e-3 &1.90e-2 &1.73e-05 &4.39e-07 &1.16e-06 \\
   \hline
40 & -                  &2.11e-05 &7.06e-06 &3.93e-3 &1.36e-3 &1.88e-05 &7.99e-11 &4.02e-10 \\
   & 3.1$^{4}$          &2.11e-05 &7.13e-06 &3.93e-3 &1.33e-3 &2.20e-06 &2.64e-22 &4.51e-21 \\
   & \textbf{7.6}$^{5}$ &2.11e-05 &2.42e-05 &3.93e-3 &3.72e-2 &2.74e-05 &9.07e-07 &2.30e-06\\
   & 15.8$^{5}$         &2.11e-05 &2.17e-05 &3.93e-3 &3.44e-2 &2.66e-05 &8.81e-07 &2.18e-06 \\
   & 20.4$^{5}$         &2.11e-05 &2.19e-05 &3.93e-3 &3.47e-2 &2.64e-05 &8.91e-07 &2.21e-06 \\
   & \textbf{32.8}$^{5}$&2.11e-05 &2.15e-05 &3.93e-3 &3.44e-2 &2.60e-05 &8.96e-07 &2.19e-06 \\
   & 81.7$^{5}$         &2.11e-05 &1.54e-05 &3.93e-3 &2.65e-2 &2.40e-05 &7.81e-07 &1.75e-06 \\
\hline
45  & -                 &2.38e-05 &7.40e-06 &4.42e-3 &5.18e-3 &2.94e-05 &1.45e-07 &2.44e-07 \\
    & 3.2$^{4}$         &2.38e-05 &7.41e-06 &4.42e-3 &1.39e-3 &3.43e-06 &3.43e-22 &9.69e-21 \\
    &\textbf{6.5}$^{5}$ &2.38e-05 &2.62e-05 &4.42e-3 &5.19e-2 &3.64e-05 &1.41e-06 &3.44e-06 \\
    & 7.8$^{1}$         &2.38e-05 &7.40e-06 &4.42e-3 &1.41e-3 &2.73e-05 &9.51e-19 &7.77e-18 \\
    & 19.5$^{5}$        &2.38e-05 &2.39e-05 &4.42e-3 &4.91e-3 &3.61e-05 &1.40e-06 &3.22e-06\\
    & 23.4$^{5}$        &2.38e-05 &2.36e-05 &4.42e-3 &4.92e-3 &3.59e-05 &1.41e-06 &3.23e-06 \\
    & \textbf{42.0}$^{5}$&2.38e-05 &2.14e-05 &4.42e-3 &4.64e-3 &3.52e-05 &1.37e-06 &3.08e-06 \\
    & 69.9$^{5}$        &2.38e-05 &1.7e-05  &4.42e-3 &3.98e-3 &3.34e-05 &1.27e-06 &2.68e-06 \\
\hline
50 & -                  &2.64e-05 &7.82e-06 &4.91e-3 &1.21e-2  &4.00e-05 &4.18e-07 &6.89e-07\\
   & \textbf{8.1}$^{5}$ &2.64e-05 &2.56e-05 &4.91e-3 &6.57e-2  &4.9e-05  &1.95e-06 &4.36e-06\\
   & 14.0$^{5}$         &2.64e-05 &2.50e-05 &4.91e-3 &6.83e-2  &4.79e-05 &2.04e-06 &4.56e-06\\
   & 21.7$^{5}$         &2.64e-05 &2.44e-05 &4.91e-3 &6.62e-2  &4.73e-05 &2.01e-06 &4.43e-06\\
   & \textbf{29.1}$^{5}$&2.64e-05 &2.32e-05 &4.91e-3 &6.26e-2  &4.71e-05 &1.95e-06 &4.22e-06\\
   & 72.3$^{5}$         &2.64e-05 &1.70e-05 &4.91e-3 &5.21e-2  &4.40e-05 &1.77e-06 &3.54e-06\\
   & 144.6$^{5}$        &2.64e-05 &1.07e-05 &4.91e-3 &3.33e-2  &4.09e-05 &1.26e-06 &2.21e-06\\
   \hline
60 & -                  &3.17e-05 &9.00e-06 &5.89e-3 &3.84e-2  &6.65e-05 &1.48e-06 &2.42e-06\\
   & 3.5$^{4}$          &3.17e-05 &8.02e-06 &5.89e-3 &1.52e-3  &7.22e-06 &6.99e-22 &5.88e-20\\
   & \textbf{7.2}$^{5}$ &3.17e-05 &8.88e-06 &5.89e-3 &3.57e-2  &6.27e-05 &1.37e-06 &2.23e-06\\
   & 14.9$^{5}$         &3.17e-05 &2.44e-05 &5.89e-3 &9.97e-2  &7.29e-05 &3.32e-06 &6.96e-06\\
   & 17.8$^{5}$         &3.17e-05 &2.42e-05 &5.89e-3 &9.88e-2  &7.28e-05 &3.31e-06 &6.89e-06\\
   & \textbf{37.0}$^{5}$&3.17e-05 &2.26e-05 &5.89e-3 &9.44e-2  &7.19e-05 &3.24e-06 &6.59e-06\\
   & 92.2$^{5}$         &3.17e-05 &1.23e-05 &5.89e-3 &6.16e-2  &6.55e-05 &2.37e-06 &4.14e-06\\
  \hline
70 & -                  &3.70e-05 &9.21e-06 &6.88e-3 &5.58e-2  &9.70e-05 &2.10e-06 &3.46e-06\\
   &\textbf{39.1}$^{5}$ &3.70e-05 &1.83e-05 &6.88e-3 &1.12e-2  &1.03e-4  &4.19e-06 &7.76e-06\\
  \hline
80 & -                  &4.23e-05 &1.49e-05 &7.86e-3 &0.14     &1.51e-4  &5.40e-06 &9.55e-06\\
   &\textbf{33.9}$^{5}$ &4.23e-05 &1.74e-05 &7.86e-3 &0.14     &1.46e-4  &5.34e-06 &9.59e-06\\
  \hline
  \end{tabular}\end{center}
\raggedright
$^{1}$ This run was terminated before the core collapse due to numerical difficulties.\\
$^{2}$ This primary star has lost such a significant amount of mass that its final state will be a white dwarf.\\
$^{3}$ The final core mass of this star is such that it is a potential electron-capture supernova.\\
$^{4}$ Terminated due to the formation of a common envelope.\\
$^{5}$ The primary star of this system was uncoupled and further evolved as a single star as the secondary overflows its Roche lobe.\\
\end{table*}
\begin{table*}[t]
\caption{Wind yields in \msun{} for the binary models for \Al{}, \Cl{}, and \Ca{} illustrating the effect of different mass loss prescription on stars near the lower mass the Wolf-Rayet limit. $M_{\rm ini}$ is the initial mass in M$_{\odot}$ and $P_{\rm ini}$ is the initial period of the binary system in days. Column 3 gives the type of wind used, where Set 1 stands for using the mass-loss prescription by \cite{NugisLamers2000} for the Wolf-Rayet star, Set 2 stands for using the mass-loss prescription by \cite{Vink2017} for helium cores smaller than 4 \msun{}, and Set 3 stands for using the same prescription for helium cores smaller than 5 \msun{}. The yields are not corrected for radioactive decay that might take place during the evolution of the star. }
\begin{center}
\begin{tabular}{ccc|cc|cc|cc}
\hline 
$M _{\rm ini} $ &  $P _{\rm ini} $&Wind &$^{26}$Al& ratio & $^{36}$Cl& ratio &$^{41}$Ca& ratio\\
(M$ _{\odot}$)& (days) & &(M$ _{\odot}$) &&(M$ _{\odot}$)& &(M$ _{\odot}$)\\
\hline
10       & 2.8      &Set 1          &3.74e-08 &-    &1.01e-07 &-        &2.44e-07&-\\
         & 4.9      &Set 1          &1.02e-07 &-    &2.05e-08 &-        &8.45e-08&-\\
         & 13.1     &Set 1          &1.01e-07 &-    &4.61e-09 &-        &2.04e-08&-\\
\hline
         & 2.8      &Set 2          &3.81e-07 &10.19&4.32e-09 &0.04     &1.53e-08&0.06\\
         & 4.9      &Set 2          &1.60e-07 &1.57 &4.28e-09 &0.21     &1.97e-08&0.23\\
         & 13.1     &Set 2          &1.29e-07 &1.28 &1.16e-12 &2.52e-4  &5.52e-12&2.71e-4\\
\hline
15       & 3.8      &Set 1          &6.04e-07 &-    &1.81e-15 &-        &6.96e-15&-\\
         & 6.7      &Set 1          &4.71e-07 &-    &3.12e-15 &-        &1.24e-14&-\\
         & 16.8     &Set 1          &4.58e-07 &-    &2.73e-15 &-        &1.09e-14&-\\
\hline
         & 3.8      &Set 2          &3.90e-07 &0.65 &1.03e-20 &5.69e-6  &1.35e-19&1.94e-5\\
         & 6.7      &Set 2          &4.69e-07 &0.99 &3.06e-15 &0.97     &1.22e-14&0.98\\
         & 16.8     &Set 2          &4.58e-07 &1    &2.73e-15 &1        &1.09e-14&1\\
\hline
         & 3.8      &Set 3          &3.90e-07 &0.65 &1.03e-20 &5.69e-6  &1.35e-19&1.94e-5\\
         & 6.7      &Set 3          &1.92e-07 &0.41 &4.40e-21 &2.43e-6  &6.61e-20&5.33e-6\\
         & 16.8     &Set 3          &1.75e-07 &0.38 &4.14e-21 &1.52e-6  &6.11e-20&5.61e-6\\
\hline
\end{tabular}\end{center}

\label{HestarYields}
\end{table*}
\end{document}